\documentclass[11pt]{article}
\usepackage{amssymb}
\usepackage{mathrsfs}
\usepackage{graphics}
\usepackage{amsmath}
\usepackage{amsmath,amsthm}
\usepackage{fullpage}
\usepackage[numbers]{natbib}
\usepackage{setspace}
\usepackage[demo]{graphicx}
\usepackage{soul}
\usepackage{graphicx}
\usepackage{xcolor}
\usepackage{url}

\usepackage{booktabs} 
\usepackage[linesnumbered,ruled,noend]{algorithm2e}

\SetAlFnt{\small}
\SetAlCapFnt{\small}
\SetAlCapNameFnt{\small}
\SetAlCapHSkip{0pt}
\IncMargin{-\parindent}

\newcommand{\equationbreak}{\text{ }}

\newcommand{\bigabs}[1]{\big\lvert #1 \big\rvert}
\newcommand{\biggabs}[1]{\bigg\lvert #1 \bigg\rvert}

\newcommand{\expbest}[1]{\mathbb{E}_{o\sim F_{\mid a^*(b)}}[#1]}

\newcommand{\expany}[1]{\mathbb{E}_{o\sim F_{\mid a}}[#1]}
\newcommand{\bigequationbreak}{\text{\space\space}}
\newcommand{\ALGone}{\operatorname{ALG1}}

\usepackage{xcolor}

\newtheorem{theorem}{Theorem}
\newtheorem*{theorem*}{Theorem}
\newtheorem{corollary}{Corollary}
\newtheorem{lemma}{Lemma}
\newtheorem*{lemma*}{Lemma}

\newtheorem{proposition}{Proposition}

\newtheorem*{observation*}{Observation}

\newtheorem{remark}[theorem]{Remark}
\newtheorem{claim}{Claim}
\newtheorem{definition}{Definition}
\newtheorem{observation}{Observation}
\newtheorem{example}{Example}

\newcommand{\sw}{\operatorname{Wel}}

\usepackage{caption}
\usepackage{xcolor}

\newcounter{matriz}

\newcounter{LP}

\setcitestyle{acmnumeric}

\begin{document}

\title{Incomplete Information VCG Contracts for Common Agency}
%
\author{ Tal Alon\thanks{Technion -- Israel Institute of Technology. Email: alontal@campus.technion.ac.il.}
\and Ron Lavi\thanks{U. of Bath, U.K., and Technion -- Israel Institute of Technology. Email: ronlavi@ie.technion.ac.il.}  
\and Elisheva S. Shamash\thanks{Technion -- Israel Institute of Technology. Email: elisheva@campus.technion.ac.il.}
\and Inbal Talgam-Cohen\thanks{Technion -- Israel Institute of Technology. Email: inbaltalgam@gmail.com.}
}

\maketitle

\begin{abstract}
 We study contract design for welfare maximization in the well-known ``common agency’’ model of Bernheim and Whinston [1986]. This model combines the challenges of coordinating multiple principals with the fundamental challenge of contract design: that principals have \emph{incomplete information} of the agent’s choice of action. Motivated by the significant social inefficiency of standard contracts for such settings (which we formally quantify using a price of anarchy/stability analysis), we investigate whether and how a recent toolbox developed for the first set of challenges under a complete-information assumption – VCG contracts [Lavi and Shamash, 2019] – can be extended to incomplete information. 

%
We define and characterize the class of ``incomplete information VCG contracts (IIVCG)’’, and show it is the unique class guaranteeing truthfulness of the principals and welfare maximization by the agent. Our results reveal an inherent tradeoff between two important properties required to ensure participation in the contract: individual rationality (for the principals) and limited liability (for the agent). We design a polynomial-time algorithm for determining whether a setting has an IIVCG contract with both properties. As our main result we design a polynomial-time ``algorithmic IIVCG’’ contract: given valuation reports from the principals it returns, if possible for the setting, a payment scheme for the agent that constitutes an IIVCG contract with all desired properties. We also give a sufficient graph-theoretic condition on the population of principals that ensures the existence of such an IIVCG contract. 

\end{abstract}

\vspace{3mm}


\onehalfspacing

\newpage

\section{Introduction}
\label{sec:intro}

{\bf Common agency.} The principal-agent model is a fundamental area of microeconomic theory, capturing natural contractual relations with important practical implications \cite{bolton2005contract,Laffont,salanie,Nobel16}. At the heart of the model is a relationship between one entity ({agent}) who acts on behalf of another ({principal}); for example, corporate management and shareholders, a freelance worker and employers, the public regulator and citizens. In all of these examples, the agent (manager/freelancer/regulator) acts on behalf of \emph{multiple} principals (shareholders/employers/citizens). The extension of the basic single-agent, single-principal setting to multiple principals mirrors the extension of the basic single-seller, single-buyer \emph{auction} setting to multiple buyers. This extension was introduced in 1986 by \citet{BernheimW86}, and is known as \emph{common agency}.
This fundamental theory has many additional applications to the study of markets, for example: a marketing agency that bids for ad space on behalf of multiple advertisers \cite{decarolis2020marketing}, a professional host on {Booking.com} who manages several properties for the owners, or platforms like Airbnb
that have characteristics of an agent representing sellers (Airbnb works to promote listings of property owners).%
\footnote{It was in fact in legal dispute in the EU whether Airbnb is legally an estate agent or not.}

\vspace{1mm}

\noindent
{\bf The agency problem.} The main challenge in the principal-agent model (with one or more principals) is the \emph{agency problem} \cite{WikiAgency21}: Actions that generate good outcomes for the principal(s) are costly for the agent, who rationally acts in his own self-interest. To align the interests of the principals with those of the agent, the principals compensate the agent depending on the outcome of his action. This compensation scheme, designed to incentivize the agent to choose a desirable action, is called a \emph{contract}.
The agency problem has two necessary ingredients (if one of these is missing, the problem is resolved): 
(i) \emph{Incomplete information} of the agent's actions; (ii) Principals are better \emph{risk bearers} than the agent. 

Incomplete information means that the principals cannot directly observe the agent's choice of action, only its resulting (stochastic) outcome. The contractual payment scheme thus depends on the outcome, which serves as a noisy indicator of the action. This crucial feature of the model is known as \emph{hidden action} or \emph{moral hazard}. For example, employers of a freelancer are not aware of how much effort he invests, but do view the outcome of his efforts and compensate him accordingly. The second crucial feature is the asymmetry between principals and agents in their attitude towards risk. The principal typically has ``deeper pockets'' and can bear the risk that the agent's actions will lead to losses. The way this is incorporated into our model follows classic works like \cite{innes1990limited, gollier1997risk,carroll2015robustness}: we focus on contracts guaranteeing \emph{limited liability (LL)} for the agent, i.e., payments in the contractual relationship flow only from the principals to the agent. Indeed, in our examples, a manager does not typically pay shareholders (even if he loses their investment), nor does a freelancer pay the employers to purchase projects.%
\footnote{Other works like \cite{BernheimW86} capture the asymmetry in risk attitudes among principal and agent by explicitly modeling the agent's risk averseness. Without LL or risk averseness, the trivial but unrealistic solution of ``buying the project’’ would apply~\cite{carroll2015robustness}.}

\subsection{Multiple Principals: Challenges and a Promising Approach}
Having multiple principals raises intriguing new challenges: 
\begin{quote}
``The principal-agent problem here is intensified: not only is there still asymmetric information between the principals and agent that can bring moral hazard, but there is also \emph{asymmetric information between the principals} themselves [... S]ince \emph{principals' interests often diverge}, they face incentives to advance their individual interests instead of the joint interests by all principals[...] As a result, introducing governance to align the interests of the principals with those of the agent is much more difficult''~\cite{WikiCommon21}.
\end{quote}
\noindent
Thus, misalignment among the principals combined with their private information (their types) requires novel contractual mechanisms. In this paper we focus on the design of \emph{welfare-maximizing} contracts.
\vspace{1mm}

\noindent
{\bf First price contracts.} The classic literature on multiple principals has thoroughly explored one class of contracts, which can be referred to as \emph{first price}. A first price contract is a simple mathematical object: it is a function~$t^{\ell}$ for every principal~$\ell$ that maps an outcome generated by the agent's action to the principal's payment. The payment is simply the principal's (reported) \emph{value} for this outcome, which can obviously be different than her true value. As we show in Theorem~\ref{thm:fp}, first price contracts can be very socially inefficient (where social efficiency is considered in equilibrium, through the standard measures of price of anarchy/stability).
\vspace{1mm}

\noindent
{\bf A solution designed for complete information.}
A possible solution arises from recent work of \citet{LaviS19}. 
Their work studies a model with \emph{complete} information about the agent’s choice of action, and thus no agency problem.
To demonstrate how different the complete and incomplete information models are, consider a single principal and an agent choosing among two actions (``shirking’’ and ``working’’), the second costing him more effort. There are two possible outcomes (``failure’’ and ``success’’), and the principal’s value for the second is higher. Shirking is more likely to produce failure than working. The precise costs, values and probabilities appear in Example 3.5, and are set up such that working maximizes the expected social welfare (the principal’s expected value for the action’s outcome, less the agent’s cost for the action).  With full information the principal can simply tell the agent to work and pay him his cost, thus maximizing welfare and extracting it fully as her expected utility. But in the incomplete information world where the agent’s action is unknown, the socially-inefficient ``null’’ contract, which pays nothing and incentivizes the agent to shirk, is preferable for the principal (see Example 3.5 for details).  

Further evidence of the difference between the two models is the state of the literature in each. For complete information, the problem of welfare maximization with multiple principals and a single agent
is largely settled by \citet{bernheim1986menu} (in a paper preceding their ``Common Agency'' paper~\cite{BernheimW86}). \citet{LaviS19} address a generalization of this problem to multiple agents (the generalization is due to \cite{prat2003games}). They instantiate an abstract concept known as \emph{contractible contracts} \cite{peters2012definable}, and introduce a novel class of \emph{VCG contracts}. In contrast, in the incomplete information world, the problem of welfare maximization with multiple principals has remained largely open even for a single agent in the three decades since the common agency model was introduced by~\cite{BernheimW86}.
Several works have considered the design of contracts with incomplete information from a computational point of view~\cite{babaioff2006combinatorial,HoSV16,KleinbergK18,KR19,roughgarden2019approximately,DRT19}; some of these works have multiple agents (e.g.,~\cite{XiaoWCTY20,AlonDPTT20}), or agent private types~\cite{GuruganeshSW20}, but none to our knowledge consider multiple principals with private types.
\vspace{1mm}

\noindent
{\bf VCG contracts and incomplete information.}
In the VGC contracts of \citet{LaviS19}, principal $\ell$’s payment for an outcome is no longer just her own (reported) value from this outcome (as in first price contracts), but is a function of the other principals’ reported values as well. 
In this sense, the payment functions in VCG contracts are reminiscent of payments in VCG \emph{auctions} \cite{Vickrey61,Clarke71,Groves73}.%
\footnote{In fact, \citet{LaviS19} note that VCG contracts generalize second price (VCG) auctions; principals in the former replace the buyers in the latter, and an agent replaces the social planner.} 
Such payments are indeed used in practice -- for example, contracts with ``price match guarantee'' clauses. Moreover, computerized platforms (such as freelancing platforms) make VCG contracts easier to coordinate. 
Such platforms typically have a central intermediary who can serve as ``mechanism designer''. E.g., in online labor markets, the market platform serves as an intermediary between an agent (freelancer) and the principals who hire him. The platform has control over pricing structures and usually treats social welfare as a first-order concern \cite[][p.~55]{roughgarden2016twenty}. Other examples include marketing agencies (multiple advertisers and a single marketing agent), corporate management (multiple managers and shareholders, and a single lower-level manager), etc.

Our goal in this work is to import ideas from VCG contracts and auctions to the incomplete information world, towards making progress on the long-standing problem of welfare maximization in common agency settings. 
%
Our first challenge is that the class of VCG contracts is undefined for incomplete information settings. After devising an appropriate generalization of the definition (Section~\ref{sub:AU-IIVG}), we uncover a more serious challenge – the adaptation of VCG contracts to incomplete information introduces negative payments. In other words, under such contracts the agent is sometimes required to pay the principal, in violation of the LL property. Our results show this is no coincidence, and ultimately address the challenge through algorithmic methods (Section~\ref{sec:algorithmic}). Our main result is a polynomial-time algorithm that implements a contract with all the desired properties, including LL, in all settings for which such a contract exists.
We believe our contribution can be a first step in a potentially rich research area of utilizing contractible contracts to address agency problems: Indeed, it is as hopeless to solve all agency problems with first price contracts as it is to solve all mechanism design problems with first price auctions. More general contract classes are called for, 
and as with auctions, the computational viewpoint comes in handy.

\subsection{Our Results}

{\bf Design objectives.} Motivated by the inefficiency of first price contracts, we seek new contracts that satisfy the following desiderata. (i) Incentives: Since principals' values are privately-known, we require that principals are incentivized to report their values truthfully in dominant strategies. We also require \emph{individual rationality}~(IR) so that principals' participation is a dominant strategy, and as explained above, also the LL property so the agent faces no risk.%
\footnote{One could also consider relaxing the LL constraint by requiring not that \emph{every} contractual payment is non-negative but rather that only the \emph{aggregate} payment from the principals to the agent is non-negative for every outcome; this does not change our results (see Remark~\ref{remark:LL}).} 
(ii) Objective: The payment schemes should incentivize the agent to choose a welfare maximizing action (we leave revenue maximization%
\footnote{The concept of ``revenue'' in a multi-principal settings may require new definitions.} 
for future work).
(iii)~Computational efficiency: Our contracts should be polynomial-time computable.

\vspace{1mm}

\noindent
{\bf Beyond first price contracts.}
We begin with two examples showing that when the principals offer \emph{first price} contracts to the agent, there can be a highly inefficient equilibrium (Example~\ref{ex:POA} shows the price of anarchy is $O(\frac{1}{n})$ for $n$ principals), and it's even possible that all equilibria are inefficient (Example~\ref{ex:POS} shows the price of stability is $\frac{1}{2}+O(\frac{1}{q-1})$ for $q$ actions).
As our first conceptual contribution, in Section~\ref{sec:Model} we formulate a game that allows us to consider more complex contract formats beyond first price: 
Every principal~$\ell\in[n]$ reports a valuation~$b^{\ell}$ mapping outcomes to values. 
(The reports can be viewed as offers made to the agent, or as ``bids'' to a centralized computerized platform.)
The agent then chooses an action which generates an outcome~$o$. A 
\emph{contract} $t=(t^1,...,t^n)$
maps the reports $b=(b^1,...,b^n)$ and the realized outcome~$o$ to a payment $t^{\ell}(b,o)$ from every principal~$\ell$ to the agent. 
\vspace{1mm}

\noindent
{\bf The IIVCG family and an impossibility.}
In Section~\ref{sec:IIVCGclass} we define the \emph{Incomplete Information VCG (IIVCG)} contract family, which is dominant strategy truthful and always induces welfare maximization. 
Dominant strategy truthfulness is a significant advantage of IIVCG over first price contracts: with the latter, a decades-old question is how the principals arrive at an equilibrium, since lacking coordination they face a bargaining-like problem of how to split the welfare.
Building upon a classic result of Holmstr\"om \cite{Holmstrom79}, we show the IIVCG family is unique (note that all results stated in this section are informal): 
\begin{theorem*}[Uniqueness of IIVCG, see Theorem~\ref{thm:uniqness}] 
Every contract $t$ such that (i) the agent is induced to maximize welfare, and (ii) the principals' dominant strategy is truthful, must belong to the IIVCG family. 
\end{theorem*}

Our definition of the IIVCG family (Definition~\ref{def:IIVCG}) specifies only the \emph{expected} payments to the agent for every action, where the expectation is over the action's stochastic outcome.
This is not enough -- a contract must determine the payments for every possible outcome realization.
We address this by characterizing the outcome-specific payments that incentivize welfare maximization on behalf of the agent: 
\begin{lemma*}[Payment characterization, see Lemma~\ref{lemma:structure}]
A contract $t=(t^1,\dots,t^\ell)$ is IIVCG if and only if principal $\ell$'s payment function $t^\ell$ can be expressed using $h^\ell(\cdot)$ and $w^\ell$, where $h^\ell(\cdot)$ is a function of the other principals' reports $b^{-\ell}$, and $\sum_{\ell}w^\ell$ is a payment vector belonging to the polytope of payments incentivizing the socially efficient action. 
\end{lemma*}
\noindent
This lemma is useful since it exactly defines our ``design space'', and in particular translates to linear constraints that are the cornerstone of our algorithmic constructions.

IIVCG contracts satisfy truthfulness and welfare maximization; what about the other desiderata of IR and LL? Whether or not these properties hold depends on the choices of $h^\ell(\cdot)$ and $w^\ell$ that determine the payments for every principal $\ell$. 
A natural starting point is to choose these to make principal~$\ell$'s expected payment equal to the externality she imposes on the others, as in VCG auctions with the Clarke pivot rule~\cite{Clarke71}, as well as VCG contracts \cite{LaviS19}. We refer to this instantiation of the IIVCG family as \emph{Auction-Inspired} contracts. In Section~\ref{sub:AU-IIVG} we show that while Auction-Inspired contracts are always IR, the LL property is not necessarily satisfied. This is no coincidence -- there is in fact an inherent clash between the two properties: 
\begin{theorem*}[LL-IR tradeoff, see Theorem~\ref{thm:notAlwaysVCG}] 
There are common agency settings for which no IIVCG contract is both LL and IR. 
\end{theorem*}
\noindent
The above theorem can apply even to settings with a single principal. This impossibility result highlights how challenging it can be to design contracts under incomplete information, and how fundamentally different this model is from the complete information one.

\vspace{1mm}

\noindent
{\bf An algorithmic approach and main results.}
In light of our impossibility result, we adopt an algorithmic approach that leverages our characterization of IIVCG contracts and uses it constructively. We give a polynomial-time algorithm (Algorithm~\ref{alg:IIVCG-domain}) that determines for a given common agent setting whether or not there is an IIVCG contract for this setting that is both LL and IR. 
If such a contract exists, we design another polynomial-time algorithm (Algorithm~\ref{alg:findIIVCG}) that implements it, i.e., upon receiving reports $b=(b^1,...,b^n)$ and an outcome~$o$, it computes $h^\ell(b^{-\ell})$ and $w^\ell$ ``on the fly'' for every principal $\ell$ in order to determine the contractual payments. 
The algorithmic basis of this approach gives us flexibility that may not exist when trying to design a contract with a closed-form formula for payments.
%


\begin{theorem*}[Main result, see Theorems~\ref{thm:alg1}-\ref{thm:alg2}] 
Consider a common agency setting.
Algorithm~\ref{alg:findIIVCG} implements a (truthful and welfare maximizing) IIVCG contract that satisfies LL and IR whenever such a contract exists for the setting; Algorithm~\ref{alg:IIVCG-domain} determines existence of such a contract.
Both algorithms run in polynomial time.
\end{theorem*}

The next observation demonstrates that IIVCG contracts can achieve welfare maximization for settings in which first price contracts fail to achieve so even approximately.

\begin{observation*}[The power of IIVCG beyond first price, see Appendix~\ref{appx:ex-social}]
Algorithm~\ref{alg:IIVCG-domain} returns ``Exists'' (i.e., there exists an LL, IR IIVCG contract) when run on Examples~\ref{ex:POA} and~\ref{ex:POS}, which have bad price of anarchy (respectively, stability) for first price contracts. 
\end{observation*}

Finally, in Section~\ref{sub:k-weighted} we give an instantiation of the IIVCG family called \emph{Weighted IIVCG}. Weighted IIVCG contracts
are always LL, and we characterize the class of settings for which they are also IR -- namely, settings that satisfy a graph-theoretic condition on the graph whose nodes are the principals.
%
This completes the picture, showing the full range within IIVCG: IR and not LL (Auction-Inspired IIVCG), LL and not IR (Weighted IIVCG), and both LL and IR whenever possible (Algorithmic IIVCG).
One avenue we do not explore in this paper but leave as an open question is contracts that guarantee LL and IR always, but only approximately achieve the optimal social welfare.


%

\section{Model and Preliminaries}
\label{sec:Model}

\noindent
\textbf{Common agency settings.} We adopt the classic model of~\citet{BernheimW86}, in which there is a single \emph{agent} who is common to $n$ \emph{principals}. 
The {agent} chooses an unobservable \emph{action} that incurs a \emph{cost} on the agent himself, and stochastically leads to an observable \emph{outcome} that benefits the principals. This is formalized by a \emph{common agency setting} as we now describe. 

Every setting is a tuple~$(\mathcal{A},\psi,\mathcal{O},\mathcal{V},F)$ as follows. $\mathcal{A}$ is a set of $q$ {actions} available to the agent. $\psi$~maps an agent's action to its {cost} (we assume for simplicity no two actions have the same cost).
$\mathcal{O}$~is a set of~$m$ {outcomes}.
Each principal~$\ell \in [n]$ has a multiparameter \emph{valuation} function~$v^\ell:\mathcal{O}\to\mathbb{R}_{\ge0}$, which maps an outcome to a non-negative value, and can also be thought of as a vector in $\mathbb{R}_{\ge 0}^m$ (valuations in our model are thus succinct objects). Valuation $v^{\ell}$ belongs to a \emph{domain} $\mathcal{V}^{\ell}$, where for simplicity we assume $\mathcal{V}^{\ell}$ is a convex, polynomially-described polytope; all our results extend to general convex domains.%
\footnote{In the general case, the convex domains are represented by polynomial-time membership oracles. Convex valuation domains appear, e.g., as the main application in \cite{Holmstrom79}.}
Valuation \emph{profile}~$v$ is the tuple $(v^1,\dots,v^n)$, and
$\mathcal{V}$~is the tuple of domains $(\mathcal{V}^{1},...,\mathcal{V}^{n})$.
$F$~is a set of distributions;
an action~$a$ induces a {probability distribution}~$F_{\mid a}$ over the outcomes (with probability $p_{o\sim F_{\mid a}}(o)$ for outcome~$o$). 
Throughout, principals are indexed by~$\ell$, actions by~$j$ and outcomes by~$i$ for consistency. Towards the computational analysis, we assume that the representation size of~$(\mathcal{A},\psi,\mathcal{O},\mathcal{V},F)$ is polynomial in~$n,q$ and $m$.
\vspace{1mm}

\noindent
\textbf{Who knows what.}
The setting~$(\mathcal{A},\psi,\mathcal{O},\mathcal{V},F)$ is assumed to be public knowledge. 
The realized outcome $i$ is publicly known, while the agent's chosen action $j$ and the principals valuations $\{v^{\ell}\}$ are privately known. This makes our model a \emph{hidden action, hidden type} model (where the type is of the principals).%
\vspace{1mm}

\noindent
\textbf{Contracts.} 
A contract profile $t=(t^{1},...,t^n)$ (\emph{contract} for short)
includes a payment rule $t^{\ell}$
for each principal ${\ell}$.
We diverge from the classic model in that the payment rules depend not only on the outcome of the agent's action, but also on the reported valuations of all principals: $t^{\ell}:\mathcal{V} \times \mathcal{O} \rightarrow \mathbb{R}$. 
Such contracts are known as \emph{contractible.}
We denote principal $\ell$'s report (a.k.a.~\emph{bid}) by $b^{\ell}\in\mathcal{V}^{\ell}$ (the report need not be truthful, i.e., $b^{\ell}$ is not necessarily $v^{\ell}$).

Taken together, the $n$ (publicly known) payment rules incentivize the agent when choosing an action. This introduces a tension among the $n$ noncooperative principals to coordinate the agent's choice of action, on top of the usual tension between each principal and the self-interested agent.
What constitutes a \emph{desirable} action depends on the principals' valuations and on the design objective~(welfare, revenue, etc.). In this work we focus on welfare, as detailed below.

\vspace{1mm}

\noindent
\textbf{The game.} Fix a common agency setting. A contract~$t$ and valuation profile $v$ define the following two-stage game. First, principals simultaneously report their bids~$b=(b^1,...,b^n)$. After perceiving the bids, the agent chooses a hidden action~$a$, which imposes a distribution~$F_{\mid a}$ over the outcomes. An outcome~$o$ is realized according to this distribution, and each principal~$\ell$ pays the agent~$t^{\ell}(b,o)$.
We assume quasi-linear utilities and risk neutrality for all parties. The utilities of the game are as follows: 
The agent's utility is the sum of payments less cost of chosen action, i.e., $\sum_{\ell \in [n]}t^{\ell}(b,o) -\psi(a)$.
Thus, the agent's action~$x^*(b)$ given bid profile~$b$ maximizes his expected utility:%
\footnote{\label{ftnt:tie-breaking}As is standard in the literature, we assume the agent tie-breaks in favor of the design objective (for an example in the context of revenue see~\cite{carroll2015robustness}). In our case the objective is the (declared) social welfare. If among the utility-maximizing actions there are several with highest (declared) welfare, we assume the agent chooses to invest maximum effort, i.e., takes the one with the maximum cost. (While we feel this kind of ``eager'' tie-breaking is in line with the standard tie-breaking assumption, our results do not rely on it and would hold for any consistent tie-breaking rule.)}
$$
x^*(b) \in \arg \max_{a \in \mathcal{A}} \left\{ \sum_{\ell \in [n]}\mathbb{E}_{o\sim F_{\mid a}}[t^{\ell}(b,o)] -\psi(a) \right\}.
$$
We assume the principals know the agent is rational and can anticipate his action $x^*(b)$ given bid profile~$b$.
As for the utility of the principals, principal $\ell$'s utility is her value less payment, i.e., $v^{\ell}(o)-t^{\ell}(b,o)$. 
\vspace{1mm}

\noindent
\textbf{Equilibrium.} 
A bid profile $b$ is in \emph{equilibrium} if no principal can strictly gain in expectation by changing her bid while holding other principals' bids fixed, and given the agent's anticipated choice of action $x^*(b)$. 
Put differently, given bid profile~$b^{-\ell}$, principal $\ell$'s equilibrium bid~$b^{\ell}$ maximizes her expected utility $\mathbb{E}_{o\sim F_{\mid x^*(b)}}[{v^{\ell}(o)]}-{\mathbb{E}_{o\sim F_{\mid x^*(b)}}[t^{\ell}(b,o)}]$ among all possible bids in $\mathcal{V}^{\ell}$, 
where $b=(b^{-\ell},b^{\ell})$, and $x^*(b)$ is the agent's anticipated action.

Consider now the incomplete information game defined by contract $t$ when the valuation profile~$v$ is not fixed. A \emph{strategy} of principal $\ell$ in this game is a mapping from every possible valuation $v^{\ell}$ to a bid $b^{\ell}$. A strategy is called \emph{dominant} if it is guaranteed to maximize the principal’s expected utility for every $v^{\ell}$ no matter what the other principals bid, i.e., for every $b^{-\ell}\in \mathcal{V}^{-\ell}$ (and still assuming the rationality of the agent).
Principal~$\ell$ is \emph{truthful} as a dominant strategy if announcing her true valuation maximizes her expected utility, i.e.,
${v^{\ell} \in \arg\max_{{b}^{\ell}\in \mathcal{V}^{\ell}} \mathbb{E}_{o\sim F_{\mid x^*(b)}}[{v^{\ell}(o) -t^{\ell}(b,o)}  ]}$ for every $v^{\ell}\in \mathcal{V}^{\ell}, b^{-\ell}\in \mathcal{V}^{-\ell}.$ We say that contract $t$ is \emph{truthful} if every principal is truthful as a dominant strategy.

\vspace{1mm}

\noindent
\textbf{Design objectives.} 
Our desiderata from a contract $t$ concern (1) incentives, (2) social welfare, and (3) computational complexity.
The first group of design objectives address incentives -- principals' and agent's willingness to participate, and principals' willingness to provide private information.
For the latter, we focus on truthful contracts, 
in which every principal's dominant strategy is to simply report her valuation. 
To ensure principals' participation in such contracts, we require that every truthful principal has non-negative expected utility no matter how the others bid -- this is known as the \emph{individual rationality (IR)} property for principals. Formally, contract~$t$ satisfies IR if $\mathbb{E}_{o \sim F_{\mid x^*(b^{-\ell},v^{\ell})}}[v^{\ell}(o)-t^{\ell}((b^{-\ell},v^{\ell}),o)]\geq 0$ for every ${\ell \in [n], b^{-\ell}\in \mathcal{V}^{-\ell},v^{\ell}\in \mathcal{V}^{\ell}}.$ 
We now turn to the agent's incentives, for which we need the following central definition: 

\begin{definition}
A contract $t$ satisfies \emph{limited liability (LL)} if every payment from a principal to the agent is non-negative: $t^{\ell}(b,o)\geq 0$ for every $\ell\in [n], b\in\mathcal{V},o\in\mathcal{O}$.
\end{definition}
\noindent The LL requirement guarantees that the agent never pays out-of-pocket. Under LL, to ensure the agent is guaranteed non-negative utility, 
we assume there exists an action with zero cost.%
\footnote{This assumption is for simplicity; the alternative of requiring IR for the agent would not change our results.}

The second design objective is social welfare. 
Given a valuation profile~$v$, the (expected) \emph{welfare} of action~$a$, denoted by $\sw^a(v)$, is the total expected value less cost, and $a^*(v)$ is a welfare-maximizing action:
\begin{equation*}
\sw^a(v)={\sum_{\ell \in [n]}\mathbb{E}_{o \sim F_{\mid a}}[v^{\ell}(o)] -\psi(a)};
\bigequationbreak
a^*(v) \in \arg\max_{a\in \mathcal{A}}{\sw^a(v)}.
\end{equation*}
We seek contracts such that for every bid profile $b$, the agent's action choice $x^*(b)$ is ``declared welfare maximizing''; we denote this by $x^*(b)=a^*(b)$.%
\footnote{If there are multiple welfare-maximizing actions then by writing $x^*(b)=a^*(b)$ we mean the agent chooses one of them. Since $x^*(b)$ is uniquely defined by the tie-breaking rule (see Footnote~\ref{ftnt:tie-breaking}), if $x^*(b)=a^*(b)$ then $a^*(b)$ is also uniquely defined.} 
Since we focus on truthful contracts, when playing according to their dominant strategies the principals report $b=v$, and the agent's choice $a^*(v)$ is indeed socially efficient. 

The last design objective is computational complexity: given a bid profile $b$ and an outcome $o$, the payments $t^{\ell}(b,o)$ should be computable in polynomial time for every $\ell\in[n]$. 
To summarize our desiderata, we seek truthful, IR and LL contracts, which incentivize the agent to maximize welfare and whose payments are computable in polynomial time.


\section{The IIVCG Family of Contracts}
\label{sec:IIVCGclass}

In this section we introduce IIVCG contracts. After giving motivation in Section~\ref{sec:FP}, we define the IIVCG family and show it is (uniquely) truthful and welfare-maximizing in Section~\ref{sub:IIVCG-def}. In Section~\ref{sub:agentsocialdesigner} we give the payment format and in Section~\ref{sub:AU-IIVG} we instantiate this format such that payments capture externalities, \`a la VCG \emph{auctions}. We show this instantiation is IR but not LL, and in Section~\ref{sub:tradeof} establish an inherent tradeoff between these two properties.

\subsection{Motivation: The Social Inefficiency of First Price Contracts}
\label{sec:FP}

A \emph{first price} contract is a contract where each principal simply pays her bid, that is,~$t^{\ell}(b,o)=b^{\ell}(o)$. As discussed in the introduction, this is the most common contract format in the literature. It satisfies the IR and LL properties. However, as we next show, it suffers from inefficiency in terms of social welfare. Since first price contracts are not truthful, we address their inefficiency through the measures of price of anarchy and price of stability.%
\footnote{The \emph{price of anarchy} is the ratio between the worst equilibrium welfare and the optimal welfare. 
The \emph{price of stability} is the ratio between the best equilibrium welfare and the optimal welfare. For details see, e.g., \cite[][Chapter 17]{AGT}.}

\begin{theorem}\label{thm:fp}
The price of anarchy of first price contracts tends to zero when the number of principals tends to infinity. The price of stability of first price contracts tends to $\frac{1}{2}$ when the number of actions tends to infinity.\footnote{We do not know whether $\frac{1}{2}$ is tight.}
\end{theorem}

The proof relies on the following examples.

\begin{example}[Price of anarchy]
\label{ex:POA}
There are three actions~$a_1,a_2,a_3$ and three outcomes~$o_1,o_2,o_3$. Action~$a_i$ deterministically yields outcome~$o_i$. The costs are: $\psi(a_1)=0,\psi(a_2)=\epsilon,\psi(a_3)=\gamma$, where $0<\epsilon<\gamma<1$. There are $n>3$ principals with the following valuation domains: $\mathcal{V}^{1}=\{(0,0,v)\mid v\geq 0\}$, $\mathcal{V}^{2}=\{(0,v,0)\mid v\geq 0\}$, and $\mathcal{V}^{\ell}=\{(v,0,0)\mid v\geq 0\}$ $\forall \ell>2$. The valuations are: $v^{1}=(0,0,1+\gamma)$, $v^{2}=(0,1+\epsilon,0)$, and $v^{\ell}=(1,0,0)$ $\forall \ell > 2$.
\end{example}

In this example, the bid profile~$b^{1}=(0,0,1+\gamma)$, $b^{2}=(0,1+\epsilon,0)$, and ${b^{{\ell}}=(0,0,0)}$ $\forall \ell > 2$ induces the agent to take action
$x^*(b)=a_{3}$, and forms an equilibrium of the game. Note that $\sw^{a_3}(v)=1$, while $\sw^{a_1}(v)=n-2$. Hence, the price of anarchy tends to zero when~$n \to \infty$. The full analysis appears in Appendix~\ref{Appx:POA}. 


\begin{example}[Price of stability]
\label{ex:POS}
There are $q\geq 2$ actions $a_1,...,a_q$ and two outcomes $o_1,o_2$. Action $a_i$ yields outcome $o_2$ with probability~$p_{o\sim F_{\mid a_i}}(o_2)=\gamma^{q-i}$ where $0<\gamma<\frac{1}{q}$.
The cost for action~$a_i$ is $\psi(a_i)={\gamma^{1-i}}-i+(i-1)(\gamma+\epsilon)$, where $0< \epsilon \leq \frac{1}{q}-\gamma$.
There is one principal with valuation domain~$\mathcal{V}^{1}=\mathbb{R}^2_{\geq q}$ and valuation $v^{1}=(q,q+{\gamma^{1-q}})$.
\end{example}

In this example, ${\sw^{a_i}(v)}=q+i-(i-1)(\gamma+\epsilon)$. Thus, the socially efficient action is $a_q$, while the action that minimizes welfare is $a_1$.
In Appendix~\ref{Appx:POS} we show that, when using a first price contract, the agent takes $a_1$ in all equilibria. The intuitive reason for this is that the principal bids to maximize her own utility rather than social welfare. Hence, the price of stability is ${\sw^{a_1}(v)}/{\sw^{a_q}(v)}$ which is at most $({q+1})/{(2q-1)}$. The last term tends to $\frac{1}{2}$ when $q\to\infty$. 
%
While the starting point of Example~\ref{ex:POS} is the ``gap setting'' construction of \cite[][p. 15]{roughgarden2019approximately}, we must modify the construction in a nontrivial way and provide a new analysis (the gap in that construction is between the optimal welfare and revenue, whereas we seek a gap between the optimal welfare and equilibrium welfare).
 
\subsection{Definition, Properties and Uniqueness of the IIVCG Family}
\label{sub:IIVCG-def}

As our main conceptual contribution, we propose to overcome the inefficiency of first price contracts by introducing the following family of contracts, which we show is (uniquely) truthful and welfare-maximizing. 

\begin{definition}
\label{def:IIVCG}
A contract~$t=(t^1,...,t^n)$ is called \emph{Incomplete Information VCG (IIVCG)} if the following two properties hold:
\begin{enumerate}
    \item {\sc Declared welfare maximization.} Given contract $t$, for every bid profile $b\in\mathcal{V}$, the agent chooses to maximize the declared social welfare, that is, $x^*(b)=a^*(b)$.\label{def:IIVCG-item:agentSocialDesigner}
    \item {\sc Expected payment characterization.} There exist functions $h^1,\dots,h^n$ where $h^{\ell}: \mathcal{V}^{-\ell} \to \mathbb{R}$, such that for every bid profile $b\in \mathcal{V}$, principal $\ell$'s expected payment $\mathbb{E}_{o \sim F_{\mid x^*(b)}}[t^{\ell}(b,o)]$ for action $x^*(b)=a^*(b)$ is equal to
    $h^{\ell}(b^{-\ell})-\sw^{a^*(b)}(b^{-\ell},\textbf{0})$ .\label{def:IIVCG-item:hs}
\end{enumerate}
\end{definition}


The first property of IIVCG contracts requires that the agent maximizes the \emph{declared} social welfare. Therefore, when principals bid truthfully, the agent takes a welfare-maximizing action, and social efficiency is achieved.
The second property characterizes the expected payment from each principal to the agent for his action $a^*(b)$. The characterization for principal $\ell$ uses $h^{\ell}$, a function of the bids of the other principals besides $\ell$, and $\sw^{a^*(b)}(b^{-\ell},\textbf{0})$, the part of the welfare from action $a^*(b)$ that is due to the other principals.%
\footnote{Technically, $\sw^{a^*(b)}(b^{-\ell},\textbf{0})$ is the welfare from $a^*(b)$ had principal $\ell$'s valuation been a zero vector, i.e., the welfare of all principals but $\ell$.}

The characterization in Definition \ref{def:IIVCG} determines only the payments in expectation (taken over the random outcome) for a certain action, not the payment rules themselves. 
We further develop the payment rules in Section \ref{sub:agentsocialdesigner}.  But Definition \ref{def:IIVCG} suffices to establish truthfulness, and thus welfare maximization, for every IIVCG contract. We also show the other direction, i.e., a sense in which this family of contracts is uniquely truthful and welfare-maximizing.

\begin{proposition}
\label{prop:truthfullVCG}
Every IIVCG contract is truthful.
\end{proposition}

\begin{proof}
Let $t$ be an IIVCG contract. Consider principal $\ell$ and fix the other principals' bids $b^{-\ell}$. We show that reporting truthfully maximizes principal $\ell$'s expected utility. The key observation is that by Property~\ref{def:IIVCG-item:hs} of Definition~\ref{def:IIVCG} of IIVCG contracts, principal~$\ell$'s expected utility 
$\expbest{v^{\ell}(o)-t^{\ell}(b,o)}$ for reporting $b^{\ell}$ is equal to $\sw^{a^*(b)}(b^{-\ell},{v}^{\ell})-h^{\ell}(b^{-\ell}).$ 
The first term is maximized when~$b^{\ell}=v^{\ell}$, because it is $\sw^a(b^{-\ell},{v}^{\ell})$ when $a=a^*(b^{-\ell},{b}^{\ell})$, and due to the fact that
$\max_a\sw^a(b^{-\ell},{v}^{\ell})=\sw^{a^*(b^{-\ell},{v}^{\ell})}(b^{-\ell},{v}^{\ell})$.
As for the second term, $h^{\ell}(b^{-\ell})$ does not depend on principal $\ell$'s report $b^{\ell}$. We conclude that the total expected utility is maximized when~$b^{\ell}=v^{\ell}$. Therefore, reporting truthfully is a dominant strategy for principal $\ell$.
\end{proof}{}

The next corollary follows immediately from Proposition~\ref{prop:truthfullVCG} and from Property~\ref{def:IIVCG-item:agentSocialDesigner} of IIVCG contracts. 

\begin{corollary}
\label{observation:maxSocialWelfareNewVCG}
In every IIVCG contract the agent maximizes social welfare.
\end{corollary}


We now show a uniqueness result for IIVCG: Whenever the agent maximizes declared welfare, IIVCG contracts are the only contracts that incentivize principals to bid truthfully as a dominant strategy. This result is reminiscent of the well-known result by \citet{Holmstrom79}, establishing uniqueness of VCG \emph{auctions} for convex domains.%
\footnote{Holstr\"om's result is a generalization of the one by \citet{GreenL77}, and in fact applies more generally to smoothly connected domains.} 
In fact, our proof makes use of a lemma from~\cite{Holmstrom79}, in addition to the convexity of the valuation domains. We give here a sketch of the proof and defer the full proof to Appendix~\ref{appx:holms}.

\begin{theorem}[Uniqueness]
\label{thm:uniqness}
A truthful contract $t$ that satisfies $x^*(b) = a^*(b)$ $\forall b\in \mathcal{V}$ is an~IIVCG contract.
\end{theorem}

\begin{proof}[Proof sketch]
To prove $t$ is IIVCG we must show existence of functions $\{h^{\ell}\}$ that satisfy Property~\ref{def:IIVCG-item:hs} in Definition~\ref{def:IIVCG} of IIVCG contracts.
We first define $h^{\ell}:\mathcal{V}\to\mathbb{R}$, i.e., as a function of the \emph{full} bid profile. The definition is as follows: $h^{\ell}(b)=\mathbb{E}_{o \sim F_{\mid a^*(b)}}[t^{\ell}(b,o)]+\sw^{a^*(b)}(b^{-\ell},\textbf{0})$ $\forall b\in \mathcal{V}$. 
In order to establish that~$t$ is an IIVCG contract, it suffices to show that $h^{\ell}(b^{-\ell},\cdot)$ is constant on~$b^{\ell}$ $\forall b^{-\ell}\in \mathcal{V}^{-\ell}$. To do so, we fix $b^{-\ell}$ and apply a key lemma from~\cite{Holmstrom79}. Roughly, this lemma states that if two functions~$f:\mathcal{V}^{\ell}\times \mathcal{V}^{\ell}\to \mathbb{R}$, $g:\mathcal{V}^{\ell}\to \mathbb{R}$ satisfy: (i) $v^{\ell}\in \arg\max_{b^{\ell}}f(b^{\ell},v^{\ell}),$ and (ii) $v^{\ell}\in \arg\max_{b^{\ell}}f(b^{\ell},v^{\ell})+g(b^{\ell})$ $\forall b^{\ell},v^{\ell}\in \mathcal{V}^{\ell}$, then $g(\cdot)$ is constant on $b^{\ell}$. Take $f(b^{\ell},v^{\ell})=\sw^{a^*(b)}(b^{-\ell},v^{\ell})$ $\forall b^{\ell},v^{\ell}\in \mathcal{V}^{\ell}$. It is immediate that (i) holds. Take $g(b^{\ell})=-h^{\ell}(b)$ $\forall b^{\ell}\in \mathcal{V}^{\ell}$. We show that (ii) also holds. Since principal $\ell$ is truthful as a dominant strategy, $v^{\ell} \in \arg\max_{b^{\ell}\in \mathcal{V}^{\ell}}\expbest{v^{\ell}(o)-t^{\ell}(b,o)}$. That is, 
$v^{\ell} \in \arg\max_{b^{\ell}\in \mathcal{V}^{\ell}}\sw^{a^*(b)}(b^{-\ell},{v}^{\ell})-h^{\ell}(b)$. Thus, according to the key lemma, $h^{\ell}(b^{-\ell},\cdot)$ is constant on~$b^{\ell}$.
\end{proof}

\subsection{A Characterization of the Outcome-Specific Payment Scheme}
\label{sub:agentsocialdesigner}

The definition of IIVCG contracts in the previous section includes a characterization of expected payments for action $a^*(b)$. In this section we complete the picture by characterizing the outcome-specific payments for all actions (Lemma~\ref{lemma:structure}). The characterization relies on Property~\ref{def:IIVCG-item:agentSocialDesigner} of IIVCG contracts -- that the agent always maximizes the declared social welfare.
Since the agent acts in his own self-interest, the payments must \emph{incentivize} the agent to choose the welfare-maximizing action. We use this fact to deduce the format of the payments. 

The characterization makes use of the following definition. For every action $a$ define 
\begin{equation*}
\mathcal{L}_{a}=\bigg\{ w \in \mathbb{R}^m_{\geq 0} \equationbreak \bigl\lvert \equationbreak a \in \arg\max_{a'}\{\mathbb{E}_{o\sim F_{\mid a'}}[w(o)] -\psi(a')\} \bigg\},
\end{equation*}
where $w(o)$ for $o\in[m]$ is the $o$th entry of vector $w$.
$\mathcal{L}_a$ can be interpreted as follows: It is the set of all non-negative payment vectors $w$ (with one payment per outcome) such that action $a$ maximizes the agent's expected utility given $w$. 
Indeed, if the principals were cooperating rather than acting each in their own self-interest, to incentivize the agent to take action~$a$ they could have jointly paid the agent~$w(o)$ for every outcome~$o$.

The following lemma uses $\mathcal{L}_{a}$ to characterize the payments: 

\begin{lemma}[Outcome-specific payment characterization]
\label{lemma:structure} 
A contract~$t$ is an IIVCG contract if and only if there exist functions~$h^{1},...,h^{n}$ where $h^{\ell}: \mathcal{V}^{-\ell} \to \mathbb{R}$, and for every~$b\in \mathcal{V}$ there exist vectors $w^1,...,w^n\in \mathbb{R}^m_{\geq 0}$ where $
\sum_{{\ell \in [n]}} w^{\ell} \in \mathcal{L}_{a^*(b)}$, such that 
$$
t^{\ell}(b,o)=h^{\ell}(b^{-\ell})-\sw^{a^*(b)}(b^{-\ell},w^{\ell}) + w^{\ell}(o) \bigequationbreak\forall{\ell \in [n], o\in \mathcal{O}}.
$$
\end{lemma}
Note that the payment format is parameterized by functions $\{h^{\ell}\}$ and vectors $\{w^{\ell}\}$, which must still be specified to get a concrete instance of an IIVCG contract, as we do in Section \ref{sub:AU-IIVG}.
The functions $\{h^{\ell}\}$ in Lemma~\ref{lemma:structure} are the same functions $\{h^{\ell}\}$ in Property~\ref{def:IIVCG-item:hs} of IIVCG contracts (Definition~\ref{def:IIVCG}).
The vectors $\{w^{\ell}\}$ are specific to a bid profile $b$. Note that for every action $a^*(b)$ that maximizes the declared welfare, $\mathcal{L}_{a^*(b)}$ is nonempty.

The proof of Lemma~\ref{lemma:structure} appears in Appendix~\ref{appx:structure}. We outline here one of the directions, i.e., that the conditions in Lemma~\ref{lemma:structure} are sufficient for contract $t$ to be IIVCG: Consider for simplicity a bid profile $b$ for which there is a single action $a^*(b)$ that maximizes the declared welfare. The agent's chosen action $x^*(b)$ maximizes his expected utility, i.e., belongs to $\arg\max_{a'\in \mathcal{A}}\sum_{\ell \in [n]}\mathbb{E}_{o\sim F_{\mid a'}}[t^{\ell}(b,o)] -\psi(a').$ Since the only term in~$t^{\ell}(b,o)$ that depends on the agent's choice of action is~$w^{\ell}(o)$, and since summing up this term over the principals gives vector $\sum_{{\ell \in [n]}} w^{\ell} \in \mathcal{L}_{a^*(b)}$, it follows that~$x^*(b)=a^*(b)$.\footnote{Recall that the agent tie-breaks in favor of the declared welfare-maximizing action $a^*(b)$.} Thus Property~\ref{def:IIVCG-item:agentSocialDesigner} of IIVCG holds. Property~\ref{def:IIVCG-item:hs} can also be verified, by observing that the expected payment is $\mathbb{E}_{o \sim F_{\mid a^*(b)}}[t^{\ell}(b,o)]=h^{\ell}(b^{-\ell})-\sw^{a^*(b)}(b^{-\ell},\textbf{0})$ $\forall b\in \mathcal{V}$. See Appendix~\ref{appx:structure} for details.


\subsection{IIVCG Instantiation: Auction-Inspired IIVCG}
\label{sub:AU-IIVG}

The payments developed in the previous section (Lemma~\ref{lemma:structure}) are parameterized by functions $\{h^{\ell}\}$ and bid-dependent vectors $\{w^{\ell}\}$. In this section we give a first instantiation of the IIVCG family of contracts by specifying these functions and vectors.
The instantiation is closely related to the VCG contracts of \cite{LaviS19} as well as to VCG \emph{auctions}, so we refer to the resulting contracts as ``Auction-Inspired IIVCG''. 

\begin{definition}
\label{def:auction-VCG}{}
Contract $t$ is an \emph{Auction-Inspired IIVCG} contract if
$$
t^{\ell}(b,o) = \sw^{a^*(b^{-\ell},\textbf{0})}(b^{-\ell},\textbf{0}) -
\sw^{a^*(b)}(b) +{{b}^{\ell}(o)}\bigequationbreak\forall {\ell \in [n]},b\in \mathcal{V}, o\in \mathcal{O}.
$$
\end{definition}

\begin{observation}
Every Auction-Inspired IIVCG contract belongs to the IIVCG family of contracts.
\end{observation}

\begin{proof}
Follows from Lemma~\ref{lemma:structure} by setting $h^{\ell}(b^{-\ell})=\sw^{a^*(b^{-\ell},\textbf{0})}(b^{-\ell},\textbf{0})$, and setting $w^{\ell} = b^{\ell}$ for bid profile $b$.
\end{proof}

Since the IIVCG family is guaranteed to be truthful (Proposition~\ref{prop:truthfullVCG}) and welfare-maximizing (Corollary~\ref{observation:maxSocialWelfareNewVCG}), our main concern in what follows is with the properties of IR and LL.

It is not hard to see that in Auction-Inspired IIVCG contracts,
the expected payment of principal~$\ell$ given bid profile~$b$ is her \emph{externality} on the other principals: the difference between their (declared) expected welfare had she not participated, i.e., $\sw^{a^*(b^{-\ell},\textbf{0})}(b^{-\ell},\textbf{0})$, and their (declared) expected welfare with her participation, i.e., 
$\sw^{a^*(b)}(\textbf{0},b^{\ell})$. This is exactly as in VCG auctions with the Clarke pivot rule. 
Interestingly, we now show that this natural instantiation of IIVCG does \emph{not} satisfy all our desiderata -- while it guarantees IR, it does not necessarily guarantee LL. 
In Section \ref{sub:tradeof} we show this is no coincidence -- there is an inherent tradeoff among IR and LL.

\begin{observation}
\label{obs:AIVCGinIIVCG}
Every Auction-Inspired IIVCG contract is IR.
\end{observation}

\begin{proof}[Proof Sketch]
It can be verified that principal $\ell$'s expected utility when she truthfully bids $b^{\ell}=v^{\ell}$ and when others bid $b^{-\ell}$ is $\max_{a\in \mathcal{A}}\sw^{a}(b^{-\ell},v^{\ell})-\max_{a\in \mathcal{A}}\sw^{a}(b^{-\ell},\textbf{0}),$ which is clearly non-negative.
\end{proof}

\begin{proposition}
\label{proposition:LL}
There exist common agency settings for which Auction-Inspired IIVCG contracts do not satisfy LL.
\end{proposition}

\begin{proof}[Proof Sketch]
Consider a common agency setting with a valuation profile such that no principal has a positive externality on the others. Assume the principals are truthful (LL should hold for any bid profile, including truthful ones). 
As mentioned above, in any Auction-Inspired IIVCG, principal $\ell$'s expected payment is her externality. The principals' expected payments are therefore all zero. The LL requirement that the agent is never paid a negative amount then means that he cannot be paid a positive amount either. 
When all payments are zero, the agent will not take any action with nonzero cost, in contradiction to the requirement that the agent maximizes welfare.
\end{proof}

A specific setting that illustrates Proposition~\ref{proposition:LL} appears in the next section (proof of Theorem~\ref{thm:notAlwaysVCG}). 


\subsection{The LL-IR Tradeoff}
\label{sub:tradeof}

As we have seen, a natural IIVCG instantiation (namely Auction-Inspired IIVCG) is IR, but does not guarantee LL. In Section~\ref{sub:k-weighted} we show a different instantiation that guarantees LL but not IR.
We now show that the clash between LL and IR is unavoidable (see Appendix~\ref{appx:trade} for the full proof):

\begin{theorem}
\label{thm:notAlwaysVCG}
There exist common agency settings for which no IIVCG contract satisfies both LL and IR, even with a single principal.
\end{theorem}


\begin{proof}[Proof Sketch]
Consider the following common agency setting: There are two actions~$a_1,a_2$ and two possible outcomes~$o_1,o_2$. The distributions imposed by the actions are $p_{o\sim F_{\mid a_1}}(o_1)=\frac{1}{2}$, $p_{o\sim F_{\mid a_1}}(o_2)=\frac{1}{2}$, $p_{o\sim F_{\mid a_2}}(o_1)=0$, $p_{o\sim F_{\mid a_2}}(o_2)=1$. 
The costs for the actions are $\psi(a_1)=0$ and $\psi(a_2)=\epsilon$. 
There is one principal with the following valuation domain~$\mathcal{V}^{1}=\mathbb{R}^{2}_{\geq 0}$. 

Suppose towards a contradiction that there exists an IIVCG contract~$t$ for this setting which satisfies IR and LL. For every valuation profile $v$, contract $t$ is truthful and incentivizes the agent to take the socially efficient action (Proposition~\ref{prop:truthfullVCG} and Corollary~\ref{observation:maxSocialWelfareNewVCG}).
Consider the valuation profile $v^{{1}}=(0,\epsilon')$, where $\frac{\epsilon'}{2}>\epsilon$. Note that $\sw^{a_1}(v)=\frac{\epsilon'}{2}$ and $\sw^{a_2}(v)=\epsilon'-\epsilon$. Thus, the socially efficient action is $a_2$. Since the agent maximizes welfare, his expected payoff from~$a_2$ is at least his expected payoff from action~$a_1$. That is,
\begin{eqnarray*}
t^{1}(v,o_2) - \psi(a_2) &\geq& \frac{1}{2}\cdot t^{1}(v,o_1)+\frac{1}{2}\cdot t^{1}(v,o_2) - \psi(a_1) \iff \\
\frac{1}{2}\cdot t^{1}(v,o_2) &\geq& \frac{1}{2}\cdot t^{1}(v,o_1) +\epsilon.
\end{eqnarray*}
This implies that
\begin{equation}
t^{{1}}(v,o_2) > t^{{1}}(v,o_1)\geq 0.\label{eq:strictly-pos} 
\end{equation}
Since~$t$ also satisfies IR, the principal's expected value $\mathbb{E}_{o \sim F_{\mid a_2}}[v^{1}(o)]=\epsilon'$ from her truthful bid is at least her expected payment $\mathbb{E}_{o \sim F_{\mid a_2}}[t^{1}(v,o)]$, which by~\eqref{eq:strictly-pos} is strictly positive. By Property~\ref{def:IIVCG-item:hs} of IIVCG contracts, $\mathbb{E}_{o \sim F_{\mid a_2}}[t^{1}(v,o)] = h^1(v^{-1})-\sw^{a^*(v)}{(v^{-1},\textbf{0})}$. Putting everything together we get:
\begin{equation}
\epsilon' \ge h^1(v^{-1})-\sw^{a_2}{(v^{-1},\textbf{0})} > 0.\label{eq:three-sides}
\end{equation}
Recall that $\sw^{a_2}{(v^{-1},\textbf{0})}=-\epsilon$, thus $\epsilon' \geq h^{{1}}(v^{-{1}})+\epsilon>0.$
However, $h^{1}(\cdot)$ is independent of $v^{1}$ and so cannot depend on $\epsilon'$. This is a contradiction.
\end{proof}

\begin{remark}
With two or more principals, Theorem~\ref{thm:notAlwaysVCG} is related to the classic result of Myerson and Satterthwaite, by which there do not exist socially-efficient and budget-balanced mechanisms~\citep{MyersonS83}. Indeed, the agent in our setting can be viewed as the mechanism designer in the classic result, and the principals in our setting can be viewed as the players. However, the impossibility result of Myerson and Satterthwaite crucially relies on the existence of two players, whereas Theorem~\ref{thm:notAlwaysVCG} applies even to settings with a single principal. 
\end{remark}

\section{Algorithmic IIVCG}
\label{sec:algorithmic}


Motivated by the LL-IR tradeoff (Theorem~\ref{thm:notAlwaysVCG}), in this section we formulate a necessary and sufficient condition on a common agency setting for the existence of an IIVCG contract that is both LL and IR (Lemma~\ref{lemma:iff-applicable}).
Our characterization is constructive, yielding an IIVCG contract that satisfies both properties whenever such a contract exists. There is a conceptual difference between this contract design and Auction-Inspired IIVCG (presented in Section~\ref{sub:AU-IIVG}) or Weighted IIVCG (presented in Section~\ref{sub:k-weighted}). In the latter classes, the contract mapping $t:\mathcal{V}\times \mathcal{O} \to \mathbb{R}^n$ is expressed using a short mathematical formula. We introduce \emph{Algorithmic IIVCG} contracts, in which~$t$ is algorithm-based:
%

\begin{definition}[Contract $\ALGone$]
For a given common agency setting, $\ALGone:\mathcal{V}\times \mathcal{O}\to \mathbb{R}^n$ is the contract that for every bid profile~$b\in \mathcal{V}$ and outcome~$o\in \mathcal{O}$, maps $(b,o)$ to Algorithm~\ref{alg:findIIVCG}'s output on~$b$ and~$o$ for this setting.  
\end{definition}

The following two theorems constitute our main results:

\begin{theorem}
\label{thm:alg1}
For every common agency setting for which an IIVCG contract that satisfies LL and IR exists, contract $\ALGone$ is such a contract. Furthermore, for every bid profile~$b\in \mathcal{V}$ and outcome~$o\in \mathcal{O}$,  Algorithm~\ref{alg:findIIVCG} computes the payment profile $\ALGone(b,o)$ in polynomial time.
\end{theorem}
\noindent
To use contract $\ALGone$, one must determine for a given setting whether there exists an IIVCG contract with both properties. Algorithm~$2$ leverages our necessary and sufficient condition and provides a procedure to do so. Formally:
\begin{theorem}\label{thm:alg2}
Given a common agency setting, Algorithm~\ref{alg:IIVCG-domain} determines in polynomial time whether there exists an IIVCG contract that is both LL and IR for this setting.
\end{theorem}
\vspace{1mm}

\noindent
{\bf Proof overview.} The proofs of both theorems are given in Sections~\ref{sub:main-pf-1} and~\ref{sub:main-pf-2}, respectively. At a high-level they proceed as follows. Given a common agency setting, we present two bounds on principals' payments:
\begin{enumerate}
    \item $m^{\ell}(b)$ $\forall b\in \mathcal{V}$, is an upper bound on principal $\ell$'s expected payment in any IIVCG contract $t$ that satisfies IR, i.e., $\expbest{t^{\ell}(b,o)}\leq m^{\ell}(b)$ $\forall \ell \in [n], b\in \mathcal{V}$ (Claim~\ref{claim:IR-iff}).
    \item  $k_{a^*(b)}$ $\forall b\in \mathcal{V}$, is a lower bound on the sum of principals' expected payments in any IIVCG contract $t$ that satisfies LL, i.e., $k_{a^*(b)}\leq \sum_{\ell \in [n]}\expbest{t^{\ell}(b,o)}$ $\forall b\in \mathcal{V}$ (Claim~\ref{claim:LL_k}).
\end{enumerate}
It follows that if there exists an IIVCG contract that is both LL and IR for the given setting, then 
\begin{equation}
k_{a^*(b)}\leq \sum_{\ell\in[n]}m^{\ell}(b) \bigequationbreak \forall b\in \mathcal{V}.\label{eq:predicate}
\end{equation}
Surprisingly, the reverse direction is also true, i.e., if the predicate in Eq.~\eqref{eq:predicate} holds, one can construct an IIVCG contract that is both LL and IR (Claim~\ref{claim:construct} and Algorithm~\ref{alg:findIIVCG}). 
Section~\ref{sub:proofALG2} shows that this predicate can be computed in polynomial time. 

\subsection{Proof of Theorem~\ref{thm:alg1}}\label{sub:proofALG1}
\label{sub:main-pf-1}

In this section we explain
Algorithm~\ref{alg:findIIVCG} and prove Theorem~\ref{thm:alg1} step by step. 
As a first step, we provide the following upper bound on principals' payments.

\begin{claim}\label{claim:IR-iff}
Consider a common agency setting. Define:
\begin{equation*}
m^{\ell}(b)= \min_{\Tilde{v}^{{\ell}}\in \mathcal{V}^{{\ell}}} \sw^{a^*(b^{-{\ell}},\Tilde{v}^{\ell})}(b^{-{\ell}},\Tilde{v}^{\ell})- \sw^{a^*(b)}(b^{-{\ell}},\textbf{0})\bigequationbreak  \forall {\ell} \in [n], {b} \in \mathcal{V}.
\end{equation*}
An IIVCG-contract $t$ satisfies IR if and only if $\equationbreak\expbest{t^{\ell}(b,o)}\leq m^{\ell}(b)$ $\forall \ell \in [n], b\in \mathcal{V}$. 
\end{claim}
\begin{proof} By definition, $t$ satisfies IR if and only if each principal's expected payment when she bids truthfully is at most her expected value. That is,
$
\mathbb{E}_{o \sim F_{\mid a^*(b^{-\ell},v^{\ell}
) }}[t^{\ell}((b^{-\ell},v^{\ell}),o)] \leq
\mathbb{E}_{o \sim F_{\mid a^*(b^{-\ell},v^{\ell}
)}}[v^{\ell}(o)]$ 
$\forall \ell \in [n], b\in \mathcal{V}, v^{\ell}\in \mathcal{V}^{\ell}.
$
By Property~\ref{def:IIVCG-item:hs} in the definition of IIVCG (Definition~\ref{def:IIVCG}), the above holds if and only if
$
h^{\ell}(b^{-\ell})-\sw^{a^*(b^{-\ell},v^{\ell}
)}(b^{-\ell},\textbf{0}) \leq
\mathbb{E}_{o \sim F_{\mid a^*(b^{-\ell},v^{\ell}
)}}[v^{\ell}(o)]$ ${\forall \ell \in [n], b\in \mathcal{V}, v^{\ell}\in \mathcal{V}^{\ell}.}$
By adding $\sw^{a^*(b^{-\ell},v^{\ell}
)}(b^{-\ell},\textbf{0})-\sw^{a^*(b)}(b^{-\ell},\textbf{0})$ to both sides of the inequality, we have that $t$ satisfies IR if and only if
$
h^{\ell}(b^{-\ell})-\sw^{a^*(b)}(b^{-\ell},\textbf{0}) \leq
\sw^{a^*(b^{-\ell},v^{\ell}
)}(b^{-\ell},v^{\ell})-\sw^{a^*(b)}(b^{-\ell},\textbf{0})$ $\forall \ell \in [n], b\in \mathcal{V}, v^{\ell}\in \mathcal{V}^{\ell}.
$
Since the left-hand side is independent of~${v}^{\ell}$, the above is equivalent to the following condition (obtained by minimizing the right-hand side over all $v^{\ell}\in \mathcal{V}^{\ell}$ and dropping the $\forall v^{\ell}\in \mathcal{V}^{\ell}$ quantifier): $
h^{\ell}(b^{-\ell})-\sw^{a^*(b)}(b^{-\ell},\textbf{0}) \leq$ $\min_{\Tilde{v}^{{\ell}}\in \mathcal{V}^{{\ell}}} \sw^{a^*(b^{-{\ell}},\Tilde{v}^{\ell})}(b^{-{\ell}},\Tilde{v}^{\ell})-\sw^{a^*(b)}(b^{-\ell},\textbf{0})$ $\forall \ell \in [n], b\in \mathcal{V}
$.
That is, by Property~\ref{def:IIVCG-item:hs} in the definition of IIVCG, and by the definition of $m^{\ell}(b)$, if and only if $\expbest{t^{\ell}(b,o)}\leq m^{\ell}(b)$ ${\forall \ell \in [n], b\in \mathcal{V}}$. \end{proof}


\begin{proposition}
\label{prop:mlb-LP}
$m^{\ell}(b)$ can be computed in polynomial time by solving the following linear program and subtracting ${F_{\mid a^*(b)}}\cdot b^{-\ell}-\psi(a^*(b))$ from its solution.
\begin{align}\label{eq:mlb-LP}
\underset{{\Tilde{v}^{\ell}\in \mathcal{V}^{\ell}, h^{\ell} \in \mathbb{R}}}{\text{minimize }} &  h^{{\ell}}\tag{LP1}\\ 
\text{ subject to } & F_{\mid a_j} \cdot (b^{-\ell}+\Tilde{v}^{\ell})-\psi(a_j) \leq h^{\ell}& \forall j\in [q], \nonumber
\end{align}
where $b^{-\ell}(o)=\sum_{\ell\neq \ell'\in [n]}b^{\ell'}(o)$ $\forall o\in \mathcal{O}$.
\end{proposition}

\begin{proof}
The linear program above is solvable in polynomial time since it consists of $O(m)$ variables: ${0 \leq h^{\ell},\Tilde{v}^{\ell}\in \mathcal{V}^{\ell}\subseteq \mathbb{R}^{m}}$; and polynomially-many constraints: $O(q)$ constraints 
in addition to the polynomially-many constraints ensuring that $v\in\mathcal{V}$ (recall that $\mathcal{V}^{\ell}$ is a convex, polynomially-described polytope for every $\ell$).
Note that $\sw^{a_j}(b^{-{\ell}},\Tilde{v}^{\ell})=F_{\mid a_j} \cdot (b^{-\ell}+\Tilde{v}^{\ell})-\psi(a_j)$. Thus, the $j$th constraint ensures that the welfare from action~$a_j$ given $(b^{-\ell},\Tilde{v}^{\ell})$ is at most $h^{\ell}$. Altogether we have that the optimal welfare $\sw^{a^*(b^{-{\ell}},\Tilde{v}^{\ell})}(b^{-{\ell}},\Tilde{v}^{\ell})\leq h^{\ell}$ at the feasible region. Minimizing $h^{\ell}$ over all $\tilde{v}^{\ell}\in\mathcal{V}^{\ell}$ gives the first term in $m^{\ell}(b)$. To complete $m^{\ell}(b)$'s computation, subtract $\sw^{a^*(b)}(b^{-{\ell}},\textbf{0})={F_{\mid a^*(b)}}\cdot b^{-\ell}-\psi(a^*(b))$ from $h^{\ell}$.
\end{proof}

\begin{claim}\label{claim:LL_k}
Consider a common agency setting. Define: $k_{a}={\min}_{w\in \mathcal{L}_{a}}\expany{w(o)}$ $\forall a$. If an IIVCG-contract $t$ satisfies LL, then $k_{a^*(b)}\leq \sum_{\ell \in [n]}\expbest{t^{\ell}(b,o)}$ $\forall b\in \mathcal{V}$.
\end{claim}

\begin{proof}
By linearity of expectation it suffices to show that $(\sum_{\ell\in [n]}t^{\ell}(b,o))_{o\in\mathcal{O}}\in\mathcal{L}_{a^*(b)}$ $\forall b\in \mathcal{V}$. 
By definition of $\mathcal{L}_{a^*(b)}$, we need to show that: (i) $\sum_{\ell\in [n]}t^{\ell}(b,o)\geq 0$ $\forall o\in \mathcal{O}$. This follows from LL. (ii) $a^*(b)\in \arg\max_{a'}\{\mathbb{E}_{o\sim F_{\mid a'}}[\sum_{\ell\in [n]}t^{\ell}(b,o)]-\psi(a')\}$. This follows from the definition of IIVCG, where the agent takes $a^*(b)$.
\end{proof}

\begin{proposition}\label{prop:k-LP}
$k_{a^*(b)}$ can be computed in polynomial time by solving the following linear program.
\begin{align}\label{eq:ka-LP}
\underset{w\in \mathbb{R}^m}{\text{minimize }} & F_{a^*(b)}\cdot w\tag{LP2}
\\ 
\text{ subject to } & F_{\mid a_j}\cdot w -\psi(a_j) \leq F_{\mid a^*(b)}\cdot w-\psi(a^*(b))& \forall j\in [q], \nonumber
\\[1ex]
& w_k\geq 0& \forall k\in [m]. \nonumber
\end{align}
\end{proposition}

\begin{proof}
The linear program above is solvable in polynomial time since it consists of $O(m)$ variables: $w\in \mathbb{R}^m$, and $O(q+m)$ constraints. The first set of constraints ensures that $a^*(b)\in \arg\max_{a'}\{\mathbb{E}_{o\sim F_{\mid a'}}{[w(o)]}-\psi(a')\}$. The second set of constrains ensures that $w\in \mathbb{R}^m_{\geq 0}$. Altogether, $w\in \mathcal{L}_{a^*(b)}$ at the feasible region. Minimizing~$F_{a^*(b)}\cdot w=\mathbb{E}_{o\sim F_{\mid a^*(b)}}{[w(o)]}$ computes $k_{a^*(b)}.$
\end{proof}

\begin{lemma}[Necessary and sufficient condition]
\label{lemma:iff-applicable}
Consider a common agency setting. There exists an IIVCG contract that satisfies LL and IR if and only if $k_{a^*(b)}\leq \sum_{\ell\in[n]}m^{\ell}(b)$ $\forall b\in \mathcal{V}$.
\end{lemma}

\begin{proof}
For the forward direction, suppose that there exists an IIVCG-contract~$t$ that satisfies both LL and IR. By claims~\ref{claim:LL_k}, \ref{claim:IR-iff}, we conclude that $k_{a^*(b)}\leq \sum_{\ell \in [n]}\expbest{t^{\ell}(b,o)} \leq  \sum_{\ell \in [n]}m^{\ell}(b)$ $\forall b\in \mathcal{V}$. The backwards direction is constructive and relies on the following claim. 
\end{proof}
\begin{claim}\label{claim:construct}
If a setting satisfies $k_{a^*(b)}\leq \sum_{\ell \in [n]}m^{\ell}(b)$ $\forall b\in \mathcal{V}$, an IIVCG contract that satisfies LL and IR may be constructed by using Lemma~\ref{lemma:structure}, and 
\begin{enumerate}
    \item $h^{\ell}({b}^{-{\ell}}) = \min_{\Tilde{v}^{{\ell}}\in \mathcal{V}^{{\ell}}}\sw^{a^*({b}^{-{\ell}},\Tilde{v}^{\ell})}({b}^{-{\ell}},\Tilde{v}^{\ell})$ $\forall \ell\in [n],{b}^{-\ell}\in \mathcal{V}^{-\ell}$.\label{item:h}
    \item For every $b\in \mathcal{V}$, take $w^b\in {\arg\min}_{w\in \mathcal{L}_{a^*({b})}}\mathbb{E}_{o \sim F_{\mid a^*({b})}}[{w(o)}]$ and choose $w^1,...,w^n$ as follows: $w^{\ell}=x^b_{\ell}\cdot w^b$ where $0\leq x^b_{\ell}$, $x^b_{\ell} \cdot k_{a^*(b)} \leq m^{\ell}(b)$ $\forall {\ell} \in [n]$, and $\sum_{\ell \in [n]}x^b_{\ell}=1$.
\end{enumerate}
\end{claim}

\begin{proof}[Proof of Claim~\ref{claim:construct}]
We first show that this contract is in IIVCG, by proving that Lemma~\ref{lemma:structure} can be applied. Note that $\sum_{\ell \in [n]}w^{\ell}=w^b \in \mathcal{L}_{a^*({b})}$ $\forall {b}\in \mathcal{V}$, and that $h^{\ell}$ is indeed independent of ${b}^{\ell}$. Thus, the constructed contract is an IIVCG contract. To show LL, we must show that
${t^{\ell}(b,o)=
h^{\ell}(b^{-\ell})-\sw^{a^*(b)}(b^{-\ell},w^{\ell}) + w^{\ell}(o) \geq 0}$ $\forall \ell \in [n],b\in \mathcal{V}, o \in \mathcal{O}.$
Since $w^{\ell}(o)$ is non-negative, the last inequality holds if $h^{\ell}(b^{-\ell})-\sw^{a^*(b)}(b^{-\ell},w^{\ell}) \geq 0$ $\forall \ell \in [n],b\in \mathcal{V}.$ By adding  $\expbest{w^{\ell}(o)}$ to both sides of the inequality, we have that the above holds if $h^{\ell}(b^{-\ell})-\sw^{a^*(b)}(b^{-\ell},\textbf{0}) \geq \expbest{w^{\ell}(o)}$ $\forall \ell \in [n],b\in \mathcal{V}.$ By our choice of $h^{\ell}(b^{-\ell})= \min_{\Tilde{v}^{{\ell}}\in \mathcal{V}^{{\ell}}}\sw^{a^*({b}^{-{\ell}},\Tilde{v}^{\ell})}({b}^{-{\ell}},\Tilde{v}^{\ell})$ 
and of
 $w^{\ell}=x^b_{\ell}\cdot {\arg\min}_{w\in \mathcal{L}_{a^*({b})}}\mathbb{E}_{o \sim F_{\mid a^*({b})}}[{w(o)}]$, this is equivalent to $m^{\ell}(b)\geq x^b_{\ell} \cdot k_{a^*(b)}$ $\forall \ell \in [n],b\in \mathcal{V}$, which holds by construction.
IR follows from Claim~\ref{claim:IR-iff}, the choice of~$h^{\ell}({b}^{-{\ell}})$, and IIVCG expected payment formula: $\mathbb{E}_{o \sim F_{\mid a^*(b)}}[t^{\ell}(b,o)]=h^{\ell}(b^{-\ell})-\sw^{a^*(b)}(b^{-\ell},\textbf{0})$.
\end{proof}

\begin{observation}\label{observation:x-LP}
The following linear program computes $x$ from Claim~\ref{claim:construct} (2) in polynomial time.
\begin{align}\label{eq:x-LP}
\underset{x^b\in \mathbb{R}^n}{\text{maximize }} &\sum_{\ell \in [n]}x_{\ell}^b\tag{LP3}
\\ 
\text{ subject to }& \sum_{\ell \in [n]}x_{\ell}^b \leq 1,\bigequationbreak x^b_{\ell}\cdot k_{a^*(b)}  \leq {m^{{\ell}}}, \bigequationbreak 0\leq x^b_{\ell} &  \forall \ell \in [n]. \nonumber
\end{align}
\end{observation}

\begin{lemma}\label{lemma:ALG1}
If a setting satisfies ${k_{a^*(b)}\leq \sum_{\ell \in [n]}m^{\ell}(b)}$ ${\forall b\in \mathcal{V}}$, Algorithm~\ref{alg:findIIVCG} computes the construction in Claim~\ref{claim:construct} in polynomial time.
\end{lemma}

\begin{proof}
The first loop computes $h^{\ell}(b^{-\ell})$ $\forall \ell\in[n]$. Line~\ref{line:compute-w-k} computes $w^b$, and line~\ref{line:compute-x} computes~$x$. The last loop computes the exact payments according to Lemma~\ref{lemma:structure}. All linear programs are solvable in polynomial time as argued above (Propositions \ref{prop:mlb-LP}, \ref{prop:k-LP} and Observation~\ref{observation:x-LP}).
\end{proof}

\begin{proof}[Proof of Theorem~\ref{thm:alg1}]
According to Lemma~\ref{lemma:iff-applicable}, Claim~\ref{claim:construct} and Lemma~\ref{lemma:ALG1}, $\ALGone$ is an IIVCG contract that satisfies LL and IR for every setting where such a contract exists. Truthfulness and social efficiency follow from the definition of IIVCG. By Lemma~\ref{lemma:ALG1}, Algorithm~\ref{alg:findIIVCG} computes the payments of $\ALGone$ in polynomial time. 
\end{proof}

\begin{remark}\label{remark:LL}
One may consider relaxing the LL constraint, by requiring that only the \emph{aggregate} payment from all principals to the agent is non-negative. The proof of Claim~\ref{claim:LL_k} shows that requiring the payments from {each individual principal} to the agent to be non-negative is without loss of generality.
\end{remark}

\begin{algorithm}[t]
\SetAlgoLined
\KwIn{Bid profile~$b\in \mathcal{V}$, outcome $o\in \mathcal{O}$ and an access to the setting $(\mathcal{A},\mathcal{O},\mathcal{V},F,\psi)$.}

\For{${\ell} \in [n]$}
{
\label{line:ms-loop-start-all}

Solve \ref{eq:mlb-LP} to find $h^{\ell}$;

$m^{\ell}\gets h^{\ell}-\{{F_{\mid a^*(b)}}\cdot b^{-\ell}-\psi(a^*(b))\}$;

}
Solve \ref{eq:ka-LP} to find $k_{a^*(b)}$ and $w^b$;\label{line:compute-w-k}

\lIf{$k_{a^*(b)}>\sum_{\ell \in [n]}m^{\ell}$}{\label{line:k-less-sum}\KwRet{\textit{Impossible}}\label{line:k-ret}}

Solve \ref{eq:x-LP} to find $x$.\label{line:compute-x}

\For{${\ell} \in [n]$}{
$t^{{\ell}}(b,o) \gets  h^{\ell}-\{F_{\mid a^*(b)}\cdot (b^{-{\ell}} + x^b_{\ell}\cdot w^b)-\psi(a^*(b))\}+x^b_{\ell}\cdot w^b(o);$\
}

\KwRet{$\{t^{\ell}(b,o)\}_{\ell\in [n]}$;}\label{line:ret-t}
\caption{Returns the payments of an IIVCG contract that satisfies IR and LL for the setting, or Impossible if no such contract exists.}\label{alg:findIIVCG}
\end{algorithm}

\subsection{Proof of Theorem~\ref{thm:alg2}}\label{sub:proofALG2}
\label{sub:main-pf-2}

In this section we prove Theorem~\ref{thm:alg2}, providing a method for determining when there exists an IIVCG contract that is both LL and IR. According to the characterization in Lemma~\ref{lemma:iff-applicable}, there exists such a contract if and only if $k_{a^*(b)}\leq \sum_{\ell \in [n]}m^{\ell}(b)$ $\forall b\in \mathcal{V}.$
A key observation is that in order to test this predicate, it suffices to test only $O(q)$ bid profiles as follows (where $q$ is the number of actions).

\begin{observation}
\label{observation:iff-ALG2}
Define $\mathcal{V}_{a}=\{{v\in \mathcal{V} \mid a = a^*(v)}\}$ $\forall a$. 
Then $k_{a^*(b)}\leq \sum_{\ell \in [n]}m^{\ell}(b)$ $\forall b\in \mathcal{V}$ if and only if 
$k_a\leq {\min}_{b \in {\mathcal{V}_a }}\{\sum_{\ell \in [n]}m^{\ell}(b)\}$ $\forall a.$
\end{observation}

\noindent
Observation~\ref{observation:iff-ALG2} follows directly from the definition of $\mathcal{V}_{a}$, and will be used in the proof of Theorem~\ref{thm:alg2}.

\begin{claim}
Consider an action~$a$. The following linear program computes ${\min}_{b \in {\mathcal{V}_a }}\sum_{\ell \in [n]}m^{\ell}(b)$ in polynomial time.%
\footnote{To set ${b}\in \mathcal{V}_a$ as a linear constraint we use the constraints of $\mathcal{L}_{a}$ (as in Algorithm~\ref{alg:findIIVCG}) with weak and strict inequalities according to the tie-breaking in $a^*(\cdot)$. A standard technique for strict inequalities in a linear program is adding a small constant that depends on the bit precision, see e.g., \cite{Linear}.} 
\begin{align}\label{LP:min-mlb}
\underset{b\in \mathcal{V}_{a},\tilde{v}\in \mathcal{V},h\in\mathbb{R}^n}{\text{minimize}} &  \sum_{\ell \in [n]}\{h^{\ell}- (F_{\mid a}\cdot  b^{-\ell}-\psi(a))\}\tag{LP4}\\
\text{subject to}\quad& F_{\mid a_j}\cdot (b^{-\ell}+\tilde{v}^{\ell})-\psi(a_j) \leq h^{\ell} & \forall j\in[q], \ell \in [n]. \nonumber
\end{align}

\end{claim}
\begin{proof}
This linear program is similar to \ref{eq:mlb-LP}, which computes the first term in $m^{\ell}(b)$, and is referred to as $h^{\ell}$. We use the same set of constraints for each principal to ensure that $\sw^{a^*(b^{-{\ell}},\Tilde{v}^{\ell})}(b^{-{\ell}},\Tilde{v}^{\ell})\leq h^{\ell}$ $\forall \ell\in [n]$ at the feasible region. However, the search space and the objective are different. Instead of computing the first term in $m^{\ell}(b)$, we subtract $F_{\mid a}\cdot b^{-\ell}-\psi(a)=F_{\mid a^*(b)}\cdot b^{-\ell}-\psi(a^*(b))$ and sum over all principals to compute $\sum_{\ell\in[n]}m^{\ell}(b)$. We can do so since $\tilde{v}^{\ell}$s are independent. Further, the search space is all $b\in \mathcal{V}_{a}$ on top of all $\tilde{v}\in \mathcal{V}$ to find $\min \sum_{\ell\in[n]}m^{\ell}(b)$.
\end{proof}

\begin{proof}[Proof of Theorem~\ref{thm:alg2}]
Recall Algorithm~\ref{alg:findIIVCG}. Note that upon receiving a bid profile $b$ and an outcome $o$, in addition to computing the payments of $\ALGone$, it computes $k_{a^*(b)}\leq \sum_{\ell\in[n]}m^{\ell}(b)$, and returns \emph{Impossible} if this predicate does not hold (see line~\ref{line:k-less-sum} in Algorithm~\ref{alg:findIIVCG}). According to Observation~\ref{observation:iff-ALG2}, given a common agency setting, it suffices to test if $k_a\leq \min_{b\in\mathcal{V}_a}\sum_{\ell\in[n]}m^{\ell}(b)$ $\forall a$ to determine whether there exists an IIVCG contract that satisfies both LL and IR. This is done in Algorithm~\ref{alg:IIVCG-domain}. For each action $a$, line~\ref{line:min-mlb} uses \ref{LP:min-mlb} to find $\arg\min_{b\in \mathcal{V}_a}\sum_{\ell\in[n]}m^{\ell}(b)$. Then, using Algorithm~\ref{alg:findIIVCG}, we test if $k_{a}\leq \sum_{\ell\in[n]}m^{\ell}(b)$. Thus, Algorithm~\ref{alg:IIVCG-domain} computes $k_a\leq \min_{b\in\mathcal{V}_a}\sum_{\ell\in[n]}m^{\ell}(b)$ $\forall a$ and determines whether a ``good'' IIVCG contract exists. 
\end{proof}

\begin{algorithm}[t]
\SetAlgoLined
\KwIn{Setting $(\mathcal{A},\mathcal{O},\mathcal{V},F,\psi)$}

\For{$a \in \mathcal{A}$}
{

Solve \ref{LP:min-mlb} to find $\arg{\min}_{b \in {\mathcal{V}_a }}\sum_{\ell \in [n]}m^{\ell}(b);$\label{line:min-mlb}

\uIf{Algorithm~\ref{alg:findIIVCG} on $b$ and arbitrary $o$ returns Impossible}{\label{line:k-leq-1}

\Return {Impossible;}}
}

\Return {Possible;}

\caption{Determines if there exists an IIVCG contract that satisfies LL and IR for a given setting.}\label{alg:IIVCG-domain}
\end{algorithm}

\begin{remark}
When the principals have no private information (i.e., their types are fixed and publicly known so $V_i$ is a singleton), there are always contracts in IIVCG that satisfy LL and IR. I.e., our Algorithm~\ref{alg:IIVCG-domain} will return ``Possible'' and our Algorithm~\ref{alg:findIIVCG} will find the best such contract for the principals. As a concrete example, one LL and IR IIVCG contract is as follows. Take $h^{\ell}=\sw^{a^*(v)}(v)$, and $w^\ell=v^{\ell}$ in Lemma~\ref{lemma:structure} of the IIVCG characterization. One can verify that a principal’s payments in this contract are exactly her values. LL holds since the principal’s values are non-negative, and IR holds since the principal’s utility is zero. We mention that this contract favors the agent’s utility over that of the principals’. At the other extreme, one can use Algorithmic IIVCG. Our LP approach to this class assures that the minimum payment required to incentivize the agent is charged from the principals, thus favoring their utility over the agent’s.
\end{remark}

\section{IIVCG Instantiation: Weighted IIVCG Contracts}
\label{sub:k-weighted}


In this section we introduce a class within IIVCG which guarantees LL but not necessarily IR, and give a condition on common agency settings (``$G$-correlation'') that suffices for the existence of IR as well.
The high-level idea is as follows.
By Proposition~\ref{proposition:LL}, in order to satisfy LL, principals' expected payment should be more than their externality on others. While increasing principals' payments may damage IR, a type of contract which we term \emph{Weighted IIVCG} succeeds in providing IR for a wide family of common agency settings. The central idea is to approximate each principal's valuation according to other principals using the concept of a \emph{correlation graph}, and use this to set the payments. Correlation graphs and $G$-correlated settings are defined in Section~\ref{sub:correlation}, and in Section~\ref{sub:weighted} we use these to define Weighted IIVCG.

\subsection{Correlation Graphs and $G$-Correlated Settings} 
\label{sub:correlation}

A correlation graph (Definition~\ref{def:corr-graph}) describes how principals' valuations are correlated. (As one interpretation, principals' valuations can be intuitively thought of as reflecting their socioeconomic status, and the correlation graph as a natural way to describe it.)

\begin{definition}\label{def:corr-graph}
A \emph{correlation graph~$G$} is a weighted directed graph with vertex set~$[n]$ and 
weights on edges~${d:[n]^2\to [0,1]}$ such that $\sum_{k\in [n]}d(k,\ell)=1$, and $d(\ell,\ell)=0$ $\forall \ell \in [n]$.
\end{definition}

\begin{figure}[h]
   \centering
    \captionsetup{justification=centering,margin=1.5cm}
    \includegraphics[scale=1,trim={0 13 0 0}]{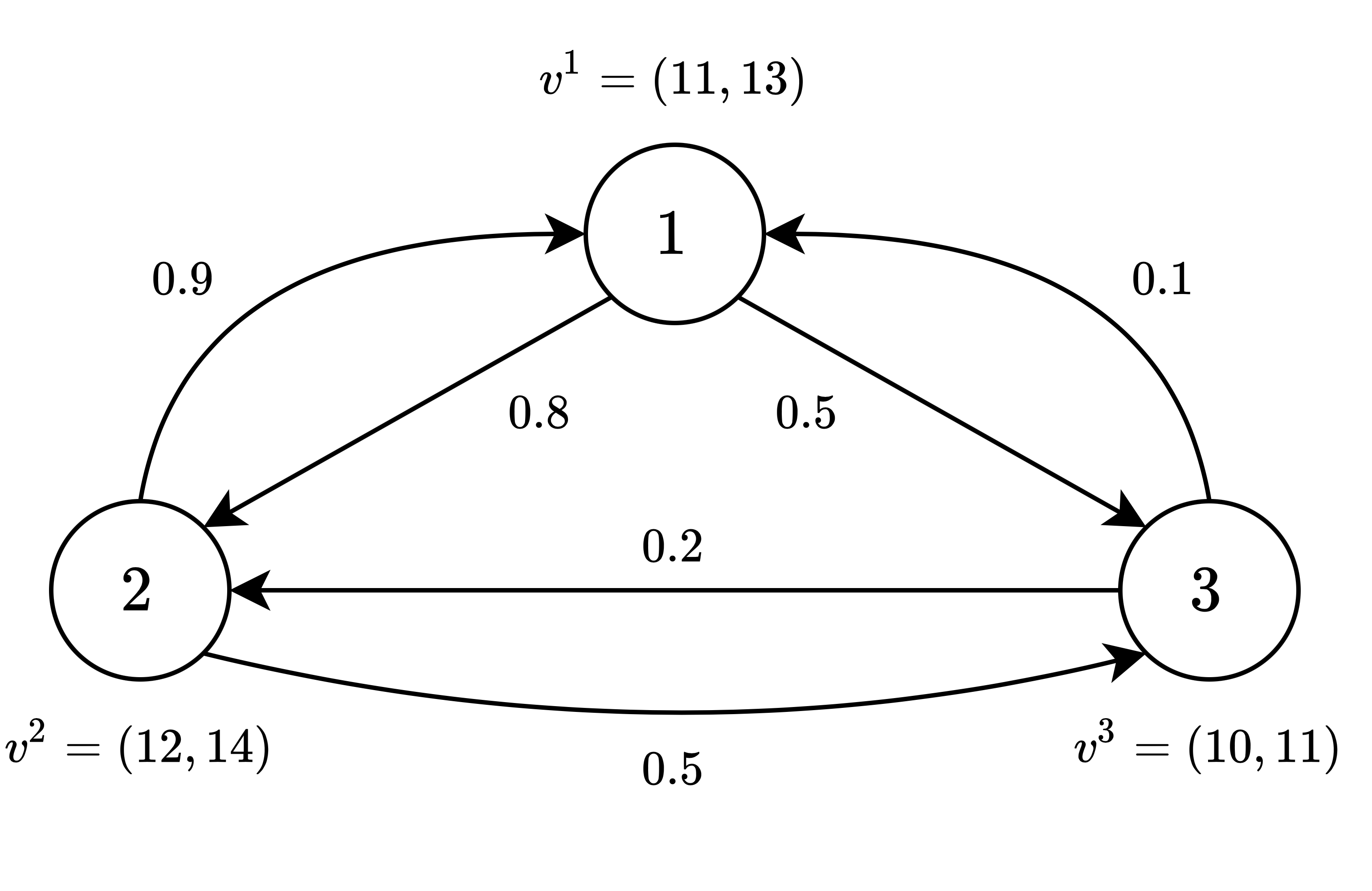}
    \caption{A possible correlation graph of the common agency setting in Example~\ref{ex:weighted}.}
    \label{fig:weighted-graph}
\end{figure}

We introduce the following ``running example'' for this section to demonstrate our definitions. Figure~\ref{fig:weighted-graph} depicts a correlation graph of the setting in the following example. 

\begin{example}\label{ex:weighted}
There are $3$ principals, $2$ actions and $2$ outcomes. The valuation domain is $\mathcal{V}^{\ell}=[10,15]^2$ $\forall \ell \in [3]$. The costs are: $\psi(a_1)=0,$ and $\psi(a_2)=1$. The probabilities over outcomes are: $p_{o\sim F_{\mid a_1}}(o_1)=1$, and $p_{o\sim F_{\mid a_2}}(o_1)=0.25$, and principals' valuations are: $v^{1}=(11,13)$, $v^{2}=(12,14)$, and $v^{3}=(10,11)$.
\end{example}

{}

We mention that a correlation graph may consist of several connected components. (In our intuitive interpretation, think of a state with multiple cities, within each city individuals are socioeconomically correlated.)

Given a correlation graph $G$, we define principals' approximate valuations, referred to as their $G$-weighted valuations, as follows.

\begin{definition}
\label{def:weightedval}
Consider a correlation graph $G$, and a valuation profile $v$. The~\emph{$G$-weighted valuation of principal~$\ell$} is: $\hat{v}^{\ell}_G=\sum_{k\in [n]}d(\ell,k)\cdot \hat{v}^k$, where $\hat{v}^{k}=v^{k}(o)-\min_{o'}v^{k}(o')$ $\forall o\in \mathcal{O},k\in[n]$.
\end{definition}

In words, the $G$-weighted valuation of principal~$\ell$ is a weighted average on other principals valuations, when their valuations are shifted downwards. We will demonstrate this for Example \ref{ex:weighted} and the correlation graph $G$ in Figure~\ref{fig:weighted-graph}: The $G$-weighted valuation of principal $1$ is $\hat{v}^{1}_G= d(1,2)\cdot \hat{v}^{2} +d(1,3)\cdot \hat{v}^{3}$. It can be verified that $\hat{v}^{2}=(0,2)$, $\hat{v}^{3}=(0,1)$, and since $d(1,2)=0.8$, and $d(1,3)=0.5$, the $G$-weighted valuation of $1$ is $\hat{v}^{1}_G=(0,1.6)+(0,0.5)=(0,2.1)$.

\begin{definition}\label{def:similargraph}
Given a correlation graph $G$, a setting is \emph{$G$-correlated} if $\equationbreak
\sw^{a^*(v^{-\ell},\hat{v}^{\ell}_G)}(v^{-\ell},\hat{v}^{\ell}_G)\leq \sw^{a^*(v)}(v)$ $\forall \ell \in [n], v\in \mathcal{V}.$
\end{definition}

That is, a setting is $G$-correlated if the welfare when principal $\ell$'s valuation is replaced with her $G$-weighted valuation, is bounded by the actual welfare (for every valuation profile). We demonstrate this condition for the setting and the valuation profile in Example~\ref{ex:weighted}, and the correlation graph in Figure~\ref{fig:weighted-graph}. Recall that $\hat{v}^{1}_G=(0,2.1)$. It can thus be verified that $\sw^{a_1}(v)=33$, $\sw^{a_2}(v)=35.75$ and that $\sw^{a_1}(v^{-\ell},\hat{v}^{\ell}_G)=22$, $\sw^{a_2}(v^{-\ell},\hat{v}^{\ell}_G)=24.825$. Thus, the welfare when principal $\ell$'s valuation is replaced with her $G$-weighted valuation is bounded by the actual welfare (for $v$). 
\vspace{1mm}

\noindent
{\bf Examples of $G$-correlated settings.} The following claims describe natural special cases for $G$-correlated settings. 

\begin{claim}\label{claim:g-corr-exp}
Consider a common agency setting. If principals share the same expected value for each action, there exists a correlation graph $G$ for which the setting is $G$-correlated.%
\footnote{While this is natural, one can show that first price contracts do well in this case: Since principals have the same expected value for each action, and principals' domains are independent, the welfare of each action is fixed. Thus, the welfare maximizing action is unified for all valuation profiles (and bid profiles). In first price contracts, the agent maximizes declared welfare. Thus the socially efficient action is always chosen by the agent.}
\end{claim}

\begin{proof}
Take $G$ as any graph on $[n]$ vertices that has a weighted in-degree and out-degree of $1$ for all vertices.\footnote{As two simple examples for such graphs, consider the graph defined by $d(\ell,\ell+1)=1$ $\forall \ell \in [n]$, where $\ell+1$ is calculated modulo $n$, and the graph is defined by $d(\ell,k)=\frac{1}{n-1}$ $\forall \ell,k \in [n]$.} Consider a valuation profile $v$ and a principal $\ell$. To satisfy the condition for $G$-correlation it suffices to have $\mathbb{E}_{o\sim F_{\mid a}}[\hat{v}^{\ell}_G(o)]\leq \mathbb{E}_{o\sim F_{\mid a}}[{v}^{\ell}(o)]$ $\forall a$. Since it implies that $\sw^{a^*(v^{-\ell},\hat{v}^{\ell}_G)}(v^{-\ell},\hat{v}^{\ell}_G)\leq \sw^{a^*(v^{-\ell},\hat{v}^{\ell}_G)}(v^{-\ell},v^{\ell})\leq \sw^{a^*(v)}(v)$. By the definition of $\hat{v}^{\ell}$ and of $\hat{v}^{k}$, $\mathbb{E}_{o\sim F_{\mid a}}[\hat{v}^{\ell}_G(o)]=\mathbb{E}_{o\sim F_{\mid a}}[\sum_{k\in [n]}d(\ell,k)\cdot\hat{v}^{k}(o)]\leq \mathbb{E}_{o\sim F_{\mid a}}[\sum_{k\in [n]}d(\ell,k)\cdot{v}^{k}(o)]$. Since principal's share the same expected valuation, $\mathbb{E}_{o\sim F_{\mid a}}[\sum_{k\in [n]}d(\ell,k)\cdot{v}^{k}(o)]=\sum_{k\in [n]}d(\ell,k)\cdot\mathbb{E}_{o\sim F_{\mid a}}[{v}^{\ell}(o)]$, and since $\sum_{k\in [n]}d(\ell,k)=1$, the last term is $\mathbb{E}_{o\sim F_{\mid a}}[{v}^{\ell}(o)]$. We conclude that $\mathbb{E}_{o\sim F_{\mid a}}[\hat{v}^{\ell}_G(o)]\leq \mathbb{E}_{o\sim F_{\mid a}}[{v}^{\ell}(o)]$ $\forall a$ as desired. This completes the proof.
\end{proof}

\begin{claim}\label{claim:range-domain}
Consider a common agency setting. If principals' valuation domain is $[a,b]^m$ where $b-a\leq a$, there exists a correlation graph $G$ for which the setting is $G$-correlated.
\end{claim}

We mention that in this case, first price contracts may fail. 

\begin{proof}
The proof proceeds as in Claim~\ref{claim:g-corr-exp}, but requires a different argument for proving that $\mathbb{E}_{o\sim F_{\mid a}}[\hat{v}^{\ell}_G(o)]\leq \mathbb{E}_{o\sim F_{\mid a}}[{v}^{\ell}(o)]$ $\forall a$. By definition, $\mathbb{E}_{o\sim F_{\mid a}}[\hat{v}^{\ell}_G(o)]=\mathbb{E}_{o\sim F_{\mid a}}[\sum_{k\in [n]}d(\ell,k)\cdot\hat{v}^{k}(o)]$. Note that since $v^{k}(o)\leq b$, and $a \leq \min_{o'\in \mathcal{O}}v^{k}(o')$, we have that $\hat{v}^{k}(o)\leq b-a\leq a$ $\forall k\in [n], o\in \mathcal{O}$. This implies that, $\mathbb{E}_{o\sim F_{\mid a}}[\sum_{k\in [n]}d(\ell,k)\cdot\hat{v}^{k}(o)]\leq \sum_{k\in [n]}d(\ell,k)\cdot a$, and since $\sum_{k\in [n]}d(\ell,k)=1$ the last term is $a$. Recall that principal's valuations are at least $a$. Thus, $a\leq \mathbb{E}_{o\sim F_{\mid a}}[{v}^{\ell}(o)]$. We conclude that $\mathbb{E}_{o\sim F_{\mid a}}[\hat{v}^{\ell}_G(o)]\leq \mathbb{E}_{o\sim F_{\mid a}}[{v}^{\ell}(o)]$ $\forall a$. This completes the proof.
\end{proof}

\subsection{$G$-Weighted IIVCG Contracts}
\label{sub:weighted}

We can now use the concepts introduced in the previous section to define Weighted IIVCG contracts and analyze their properties.

\begin{definition}\label{def:weighted-VCG}{}
Consider a correlation graph~$G$. A \emph{$G$-Weighted IIVCG Contract} is given by $$
t^{{\ell}}(b,o) = \sw^{a^*(b^{-\ell},\hat{b}^{\ell}_G)}(b^{-\ell},\hat{b}^{\ell}_G)-\sw^{a^*(b)}(b^{-\ell},\hat{b}^{\ell}_G)+\hat{b}^{\ell}_G(o)\bigequationbreak \forall \ell\in [n], b\in \mathcal{V}, o\in \mathcal{O}.
$$
\end{definition}{}

\begin{proposition}
Consider a correlation graph $G$. Every $G$-Weighted IIVCG contract is in IIVCG.
\end{proposition}
\begin{proof}
The proof relies on Lemma~\ref{lemma:structure} when $h^{{\ell}}(b^{-{{\ell}}})=\sw^{a^*(b^{-\ell},\hat{b}^{\ell}_G)}(b^{-\ell},\hat{b}^{\ell}_G)$ and $w^{{\ell}}=\hat{b}^{\ell}_G(o)$. We prove that the conditions of Lemma~\ref{lemma:structure} are met. First, $h^{{\ell}}$ is independent of $b^{\ell}$, since $\hat{b}^{\ell}_G$ is independent of $b^{\ell}$ ($d(\ell,\ell)=0$ $\forall \ell\in[n]$). Second, we show that $\sum_{\ell \in [n]}w^{\ell}\in \mathcal{L}_{a^*(b)}$. To do so, we establish that $\sum_{\ell \in [n]}w^{{\ell}}=\sum_{\ell\in [n]} \hat{b}^{\ell}$ using the following chain of equalities:
$$
\sum_{\ell \in [n]}w^{{\ell}}\stackrel{(1)}{=}\sum_{\ell \in [n]}\sum_{k\in [n]}d(\ell,k) \hat{b}^k\stackrel{(2)}{=}\sum_{k\in [n]}\sum_{\ell \in [n]}d(\ell,k) \hat{b}^k\stackrel{(3)}{=}\sum_{k\in [n]} \hat{b}^k.
$$ 
Equality (1) follows from the choice of $w^{\ell}$s above, and the definition of $\hat{b}^{\ell}_G$ (Definition~\ref{def:weightedval}). Equality (2) changes the order of summands, and Equality (3) follows from the fact that ${\sum_{k\in [n]}d(k,\ell)=1}$ ${\forall \ell \in [n]}$ (Definition~\ref{def:similargraph}). Thus, $\sum_{\ell \in [n]}w^{{\ell}}=\sum_{\ell\in [n]} \hat{b}^{\ell}$, and it suffices to show that $\sum_{\ell\in [n]} \hat{b}^{\ell} \in \mathcal{L}_{a^*(b)}$. To do so, we show that $\sum_{\ell\in [n]} \hat{b}^{\ell}\in\mathbb{R}_{\geq 0}^m$ and that $a^*(b)\in \arg\max_{a'}\{\mathbb{E}_{o\sim F_{\mid a'}}[\sum_{{\ell}\in [n]} \hat{b}^{\ell}(o)] -\psi(a')\}$. First, by definition ${\hat{b}^{\ell}(o)\geq 0}$ ${\forall o\in \mathcal{O},\ell\in[n]}$. Thus, $\sum_{\ell\in [n]} \hat{b}^{\ell}\in\mathbb{R}_{\geq 0}^m$. Second, by definition of $a^*(b)$, it holds that $a^*(b)\in \arg\max_{a'}\{\mathbb{E}_{o\sim F_{\mid a'}}[\sum_{\ell\in [n]} {b}^{\ell}(o)] -\psi(a')\}$. Thus $a^*(b)\in \arg\max_{a'}\{\mathbb{E}_{o\sim F_{\mid a'}}[\sum_{\ell\in [n]} {b}^{\ell}(o)-y] -\psi(a')\}$ $\forall y\in \mathbb{R}$, and specifically for $y=\sum_{{\ell}\in[n]}\min_{o\in \mathcal{O}}b^{\ell}(o).$ We conclude that $a^*(b)\in \arg\max_{a'}\{\mathbb{E}_{o\sim F_{\mid a'}}[\sum_{{\ell}\in [n]} \hat{b}^{\ell}(o)] -\psi(a')\}$. This completes the proof.
\end{proof}

\begin{theorem}\label{prop:KweightedinIIVCG}
Consider a correlation graph $G$. Every $G$-Weighted IIVCG contract satisfies LL. Furthermore, IR is satisfied if and only if the setting is $G$-correlated.
\end{theorem}

\begin{proof}
Truthfulness and social efficiency are immediate since $G$-Weighted IIVCG is in IIVCG. LL follows from the fact that the first term in the payment formula ,$\sw^{a^*(b^{-\ell},\hat{b}^{\ell}_G)}(b^{-\ell},\hat{b}^{\ell}_G)$, is (weakly) higher than the second term, $\sw^{a^*(b)}(b^{-\ell},\hat{b}^{\ell}_G)$, and since $\hat{b}^{\ell}_G(o)\geq 0$ $\forall o$. We prove IR if and only if the setting is $G$-correlated. By definition, IR holds if and only if each principal $\ell$'s expected payment when she bids truthfully is at most her expected value. That is, by the definition of $G$-Weighted IIVCG IR holds, if and only if ${\sw^{a^*(b^{-\ell},\hat{b}^{\ell}_G)}(b^{-\ell},\hat{b}^{\ell}_G)-\sw^{a^*(b^{-\ell},v^{\ell})}(b^{-\ell},\textbf{0})\leq \mathbb{E}_{o\sim F_{\mid a^*(b^{-\ell},v^{\ell})}}[v^{\ell}(o)]}$ $\forall b^{-\ell}\in \mathcal{V}^{-\ell}, v^{\ell}\in\mathcal{V}^{\ell},\ell\in[n]$. By adding $\sw^{a^*(b^{-\ell},v^{\ell})}(b^{-\ell},\textbf{0})$ to both sides of the inequality, we get ${\sw^{a^*(b^{-\ell},\hat{b}^{\ell}_G)}(b^{-\ell},\hat{b}^{\ell}_G)\leq \sw^{a^*(b^{-\ell},v^{\ell})}(b^{-\ell},v^{\ell})}$ ${\forall b^{-\ell}\in \mathcal{V}^{-\ell}, v^{\ell}\in\mathcal{V}^{\ell},\ell\in[n]}$. Denoting $b=(b^{-\ell},v^{\ell})$, this condition is equivalent to Definition~\ref{def:similargraph} of the $G$-correlated setting. This completes the proof.
\end{proof}


\bibliographystyle{ACM-Reference-Format}
\bibliography{bibliography}


\appendix
\section{Social Inefficiency of First Price Contracts}

\subsection{Price of Anarchy: Analysis of Example~\ref{ex:POA}}
\label{Appx:POA}

In this section we prove that in Example~\ref{ex:POA}, action $a_{3}$ and the bid profile~$b^{1}=(0,0,1+\gamma)$, $b^{2}=(0,1+\epsilon,0)$ and
${b^{{\ell}}=(0,0,0)}$ $\forall \ell > 2$ constitute an equilibrium.
To do so, we show that the agent maximizes his utility, and that each principal cannot benefit from deviating and switching her bid. Note that $\sw^{a_3}(v)=1$, while the action which maximizes social welfare is $a_1$ with $\sw^{a_1}(v)=n-2>1$.
Therefore, the price of anarchy is
$ {\sw^{a_3}(v)}/{\sw^{a_1}(v)}={1}/{(n-2)}$,
which tends to zero when $n \to \infty$.

For the agent, the payoff from taking action $a_3$ or action $a_2$ is $1$, while the payoff from action $a_1$ is $0$. Recall that the tie-breaking rule is in favor of the declared welfare and then the cost (see footnote~\ref{ftnt:tie-breaking}). Since $a_2$ and $a_3$ have the same declared welfare the agent takes action $a_3$ which incurs more cost. Therefore, $x^*(b)=a_3$. Principal $1$ cannot raise her utility by changing the bid for~$o_3$. If she raises her bid, the agent still takes $a_3$ and her utility will strictly decrease since she pays more to the agent. If principal~$1$ lowers her bid, then the agent takes $a_2$, and principal~$1$'s utility  will not vary. Principal $2$ cannot raise her utility by changing the bid for~$o_2$. If she raises her bid, the agent takes $a_2$ and her utility will strictly decrease since she pays more then her value to the agent. If principal~$2$ lowers her bid, then the agent still takes $a_3$, and principal~$2$'s utility does not vary. As for principal~$\ell>2$, she cannot raise her utility by announcing a different bid for~$o_{1}$. Assume she announces a positive bid $\tilde{b}^{\ell}(o_{1})>0$ and consider two cases: (i) If $\tilde{b}^{\ell}(o_{1})\leq 1$, the agent takes $a_3$ and the utility of principal~$\ell$ does not vary.
(ii) If $\tilde{b}^{\ell}(o_3)> 1$, the agent takes $a_1$, yet the utility of principal~$\ell$ becomes negative. We have shown that neither the agent nor the principals may benefit from deviation. Hence, action $a_3$ and the bid profile $b$ constitute an equilibrium. 

\subsection{Price of Stability: Analysis of Example~\ref{ex:POS}}
\label{Appx:POS}

In this section we prove that in Example~\ref{ex:POS}, the agent takes $a_1$ in all equilibria. The proof relies on the following Claim.
\begin{claim}\label{claim:equilibrium}
Consider Example~\ref{ex:POS}. For every bid profile that incentivizes the agent to take~$a_i$ for~$i>1$, the principal's expected utility is at most $1-\epsilon/(1-\gamma).$
\end{claim}

Observe that the smallest possible bid profile~${b}^1=(q,q)$ incentivizes the agent to take action~$a_1$. Also, the principal's expected utility given ${b}^1$ is $1>1-\epsilon/(1-\gamma)$. Thus, there is no equilibrium where the agent takes $a_i$ for $i>1$.

\begin{proof}[Proof of Claim~\ref{claim:equilibrium}]
Let $b^1$ be a bid profile that incentivizes $a_i$ for $i>1$. It must hold that $\mathbb{E}_{o\sim F_{\mid a_i}}[b^1(o)]-\psi(a_{i})\geq \mathbb{E}_{o\sim F_{\mid a_{i-1}}}[b^1(o)]-\psi(a_{i-1})$. That is, 
$
(\gamma^{q-i}-\gamma^{q-(i-1)})(b^1(o_2)-b^1(o_1))\geq \gamma^{1-i}(1-\gamma)-1+(\gamma+\epsilon). $ 
Since $b^1(o_1)$ must be at least $q$, it follows from the last inequality that $b^1(o_2)\geq q+ \gamma^{1-q}-\gamma^{i-q}+{\epsilon}/{(\gamma^{q-i}-\gamma^{q-(i-1)})}.$ Using this, the following chain of inequalities shows that the principal's expected utility is at most $1-\epsilon/(1-\gamma)$:
\begin{eqnarray*}
&&\mathbb{E}_{o\sim F_{a_i}}[v^1(o)-t^1(b,o)]\\
&=&\gamma^{q-i}(v^1(o_2)-b^1(o_2))
+(1-\gamma^{q-i})(v^1(o_1)-b^1(o_1))\\
&\leq&\gamma^{q-i}(v^1(o_2)-q-\gamma^{1-q}+\gamma^{i-q}-{\epsilon}/{(\gamma^{q-i}-\gamma^{q-(i-1)})})
+(1-\gamma^{q-i})(v^1(o_1)-q)\\
&=&\gamma^{q-i}(\gamma^{i-q}-{\epsilon}/{(\gamma^{q-i}-\gamma^{q-(i-1)})})\\
&=&1-{\epsilon}/{(1-\gamma)}.
\end{eqnarray*}
\end{proof}

\subsection{Existence of IIVCG for Examples \ref{ex:POA} and \ref{ex:POS}}
\label{appx:ex-social}

\begin{proposition}
There exists an IIVCG contract for the setting in Example~\ref{ex:POA}.
\end{proposition} 

\begin{proof}
We must show that $k_{a^*(b)}\leq \sum_{\ell \in [n]}m^{\ell}(b)$ $\forall b\in \mathcal{V}.$
We use the following case analysis:
\begin{enumerate}
    \item $a^*(b)=a_1$. Note that $k_{a_1}=0$, and that ${m^{\ell}(b)\geq 0}$ ${\forall \ell\in[n],b\in \mathcal{V}}$. Thus, $k_{a^*(b)}\leq \sum_{\ell\in [n]}m^{\ell}(b)$.
    \item $a^*(b)=a_2$. Note that $k_{a_2}=\epsilon$ by using $w=(0,\epsilon,0)$. Further, since $a^*(b)=a_2$, and by the definition of the  domain, $\sw^{a^*(b)}(b)=b^2(o_2)-\epsilon$. Thus, $\sw^{a^*(b)}(b^{-{1}},\textbf{0})=\sw^{a^*(b)}(b^{-{\ell}},\textbf{0})=b^2(o_2)-\epsilon$ $\forall \ell>2$, and $\min_{\Tilde{v}^{{1}}\in \mathcal{V}^{{1}}} \sw^{a^*(b^{-{1}},\Tilde{v}^{1})}(b^{-{1}},\Tilde{v}^{1}),\min_{\Tilde{v}^{{\ell}}\in \mathcal{V}^{{\ell}}} \sw^{a^*(b^{-{\ell}},\Tilde{v}^{\ell})}(b^{-{\ell}},\Tilde{v}^{\ell}) \geq b^2(o_2)-\epsilon$ $\forall \ell \in [n]$. This implies that $m^{1}(b)=m^{\ell}(b)=0$ $\forall \ell>2$. Further, $\sw^{a^*(b)}(b^{-{2}},\textbf{0})=-\epsilon.$ Thus, $m^2(b)\geq \epsilon.$ It follows that $k_{a^*(b)}\leq \sum_{\ell\in[n]}m^{\ell}(b)$. 
    \item  $a^*(b)=a_3$. Note that $k_{a_3}=\gamma$ by using $w=(0,0,\gamma)$. Further, since $a^*(b)=a_3$, and by the definition of the domain, $\sw^{a^*(b)}(b)=b^1(o_3)-\gamma$. Thus, $\sw^{a^*(b)}(b^{-{2}},\textbf{0})=\sw^{a^*(b)}(b^{-{\ell}},\textbf{0})=b^1(o_3)-\gamma$ $\forall \ell>2$, and $\min_{\Tilde{v}^{{2}}\in \mathcal{V}^{{2}}} \sw^{a^*(b^{-{2}},\Tilde{v}^{2})}(b^{-{2}},\Tilde{v}^{2}),\min_{\Tilde{v}^{{\ell}}\in \mathcal{V}^{{\ell}}} \sw^{a^*(b^{-{\ell}},\Tilde{v}^{\ell})}(b^{-{\ell}},\Tilde{v}^{\ell}) \geq b^1(o_3)-\gamma$ $\forall \ell \in [n]$. This implies that $m^{2}(b)=m^{\ell}(b)=0$ $\forall \ell>2$. Further, $\sw^{a^*(b)}(b^{-{1}},\textbf{0})=-\gamma.$ Thus, $m^1(b)\geq \gamma.$ It follows that $k_{a^*(b)}\leq \sum_{\ell\in[n]}m^{\ell}(b)$.  
\end{enumerate}
This completes the proof.
\end{proof}

\begin{proposition}\label{claim:applicable}
There exists an IIVCG contract for the setting in Example~\ref{ex:POS}.
\end{proposition}

To prove this, we first state the following claims.

\begin{claim}
\label{proposition:k_a}
$k_{a_j}=\gamma^{1-j}-1+\epsilon/(1-\gamma)$ $\forall j\in [q]$.
\end{claim}

\begin{claim}
\label{proposition:mlb}
$m^{\ell}(b)=q+\psi(a^*(b))$ $\forall b\in \mathcal{V}$.
\end{claim}

\begin{proof}[Proof of Proposition~\ref{claim:applicable}]
As shown in Section~\ref{sub:main-pf-1}, the setting is applicable if $k_{a^*(b)}\leq m^{1}(b)$ $\forall b\in \mathcal{V}$. Equivalently, according to Claims~\ref{proposition:k_a} and \ref{proposition:mlb}, the setting is applicable if $\gamma^{1-j}-1+\epsilon/(1-\gamma) \leq q+{\gamma^{1-j}}-j+(j-1)(\gamma+\epsilon),$ where $a_j=a^*(b)$, for all $b\in \mathcal{V}$. 
Note that $(j-1)(1-\gamma-\epsilon)+\epsilon/(1-\gamma)\leq q$ for all $j\in [q]$, since the left-hand side is at most $(q-1)(1-\gamma-\epsilon)+1$ which, in turn, is at most~$q$. This completes the proof.
\end{proof}
\begin{proof}[Proof of Claim~\ref{proposition:k_a}]
Recall that $k_{a_j}={\min}_{w\in \mathcal{L}_{a}}\mathbb{E}_{o\sim F_{\mid a_j}}[{w(o)}]$. We show that one $w$ which minimizes $\mathbb{E}_{o\sim F_{\mid a_j}}[{w(o)}]$ is
$w=(0,\gamma^{1-q}-\gamma^{j-q}+\gamma^{j-q}\cdot \epsilon  /(1-\gamma))$. Thus, $k_{a_j}=\gamma^{1-j}-1+\epsilon/(1-\gamma)$.

First, we prove that $w\in \mathcal{L}_{a_j}$.
To do so, we must show that $a_j \in \arg\max_{a'\in \mathcal{A}}\mathbb{E}_{o\sim F_{\mid a'}}[w(o)] -\psi(a')
$. Equivalently, that $\mathbb{E}_{o\sim F_{\mid a_j}}[w(o)] -\psi(a_j)\geq \mathbb{E}_{o\sim F_{\mid a_{j'}}}[w(o)] -\psi(a_{j'})$ $\forall j'\in [q]$. That is, $(\gamma^{q-j}-\gamma^{q-j'})(w_2-w_1)\geq \psi(a_j)-\psi(a_{j'})
$ $\forall j'\in [q]$. Replacing $w$ and $\psi$, the last inequality holds if 
$$
(\gamma^{q-j}-\gamma^{q-j'})(\gamma^{1-q}-\gamma^{j-q}+ \gamma^{j-q}\cdot \epsilon /(1-\gamma))\geq {\gamma^{1-j}}-{\gamma^{1-j'}}-(j-j')(1-\gamma-\epsilon)\text{\space}\forall j'\in [q].
$$ 
Reorganizing both sides, this is equal to 
$$
\gamma^{1-j}-\gamma^{1-j'}+(\gamma^{q-j}-\gamma^{q-j'})(-\gamma^{j-q}+ \gamma^{j-q}\cdot \epsilon /(1-\gamma))\geq {\gamma^{1-j}}-{\gamma^{1-j'}}-(j-j')(1-\gamma-\epsilon)\text{\space}\forall j'\in [q].
$$
Which, by subtracting  ${\gamma^{1-j}}-{\gamma^{1-j'}}$ from both sides, holds if
\begin{equation}\label{eq:prpo-k_a}
(\gamma^{q-j}-\gamma^{q-j'})(-\gamma^{j-q}+ \gamma^{j-q}\cdot \epsilon /(1-\gamma))\geq -(j-j')(1-\gamma-\epsilon)\text{\space}\forall j'\in [q].
\end{equation}
Observe the left-hand side in~\eqref{eq:prpo-k_a}, by dividing the first term by $\gamma^{q-j}$ and multiplying the second term by $\gamma^{q-j}$, it is equal to $(1-\gamma^{j-j'})(-1+\epsilon /(1-\gamma))=-(1-\gamma^{j-j'})(1-\gamma-\epsilon )/(1-\gamma)$. Then, by dividing both sides in~\eqref{eq:prpo-k_a} by $-(1-\gamma-\epsilon)$, we conclude that $w\in \mathcal{L}_{a_j}$ if
\begin{equation}\label{eq:prpo-k_a-2}
(1-\gamma^{j-j'})/(1-\gamma) \leq (j-j')\text{\space}\forall j'\in [q].
\end{equation}
The following case analysis shows the above inequality holds:
\begin{enumerate}
    \item For $j-j'=0$,$j-j'=1$, this is immediate. 
    \item  For $j-j'\geq 2$, \eqref{eq:prpo-k_a-2} holds since $ 1-\gamma^{j-j'} \leq 1 \leq 2\cdot (1-\frac{1}{q}) \leq 2\cdot (1-\gamma) \leq (j-j')(1-\gamma).$
    \item For $0>j-j'$, \eqref{eq:prpo-k_a-2} holds since $1-\gamma^{j'-j}=-(\frac{1}{\gamma}^{j'-j}-1) \leq -(2^{j'-j}-1) \leq -(j'-j) \leq (j-j')(1-\gamma)$, where the first inequality follows from the fact that $\gamma\leq \frac{1}{q}\leq \frac{1}{2}$.
\end{enumerate}
We have shown that $w\in \mathcal{L}_{a_j}$. It is now left to show that $w$ minimizes $\mathbb{E}_{o\sim F_{\mid a_j}}[{w(o)}]$. Observe that for every ${w'\in \mathcal{L}_{a}}$, it must hold that $\mathbb{E}_{o\sim F_{\mid a_j}}[{w'(o)}] \geq \mathbb{E}_{o\sim F_{\mid a_{j-1}}}[{w'(o)}]$. That is,
$(\gamma^{q-j}-\gamma^{q-j+1})(w'_2-w'_1)\geq {\gamma^{1-j}}-{\gamma^{2-j}}-(1-\gamma-\epsilon)
$. By reorganizing both sides, the last inequality equals to $\gamma^{q-j}(1-\gamma)(w'_2-w'_1)\geq ({\gamma^{1-j}}-1)(1-{\gamma})+\epsilon$. Dividing both sides by $\gamma^{q-j}(1-\gamma)$, and since $w'_1\geq 0$, it holds that $w'_2\geq \gamma^{1-q}-\gamma^{j-q}+\epsilon/(\gamma^{q-j}-\gamma^{q-(j-1)})$. Since $w$ meets these lower bounds, it minimizes $\mathbb{E}_{o\sim F_{\mid a_j}}[{w(o)}]$.  This completes the proof. 
\end{proof}

\begin{proof}[Proof of Claim~\ref{proposition:mlb}] Recall that $m^{1}(b)= \min_{\Tilde{v}^{{1}}\in \mathcal{V}^{{1}}} \sw^{a^*(b^{-{1}},\Tilde{v}^{1})}(b^{-{1}},\Tilde{v}^{1})- \sw^{a^*(b)}(b^{-{1}},\textbf{0}).$
Note that $\sw^{a^*(b)}(b^{-{1}},\textbf{0})=-\psi(a^*(b))$. Further, $\min_{\Tilde{v}^{{1}}\in \mathcal{V}^{{1}}} \sw^{a^*(b^{-{1}},\Tilde{v}^{1})}(b^{-{1}},\Tilde{v}^{1})=q$, when $\tilde{v}^{1}=(q,q)$: Since for every $v^1\neq \tilde{v}^1\in \mathcal{V}^1$ it holds that ${\Tilde{v}^1\leq {v}^{1}}$ (element wise). Thus, $\sw^{a^*(b^{-1},\Tilde{v}^{1})}(b^{-1},\Tilde{v}^{1})$ $\leq \sw^{a^*(b^{-1},\Tilde{v}^{1})}(b^{-1},{v}^{1})\leq  \sw^{a^*(b^{-1},{v}^{1})}(b^{-1},{v}^{1}).$ This completes the proof.
\end{proof}
To complete the picture, one IIVCG contract for Example~\ref{ex:POS} may be contracted using $h^1(b^{-1})=q$ and $w=(0,\gamma^{1-q}-\gamma^{j-q}+\gamma^{j-q}\cdot \epsilon  /(1-\gamma))$ where $a_j=a^*(b)$ $\forall b\in \mathcal{V}$. That is,
\begin{eqnarray*}
t^1(b,o_1)&=&q-\{(j-1)(1-\gamma-\epsilon)+\epsilon/(1-\gamma)\}.\\
t^1(b,o_2)&=&q-\{(j-1)(1-\gamma-\epsilon)+\epsilon/(1-\gamma)\}+\gamma^{1-q}-\gamma^{j-q}+\gamma^{j-q}\cdot \epsilon  /(1-\gamma).
\end{eqnarray*}
This contract satisfies limited liability because $q\geq (q-1)(1-\gamma-\epsilon)+1\geq (j-1)(1-\gamma-\epsilon)+\epsilon/(1-\gamma),$ and
$\gamma^{j-q}-\gamma^{j-q}\cdot \epsilon  /(1-\gamma)=\gamma^{1-q}\cdot \gamma^{j-1}\cdot(1- \epsilon  /(1-\gamma))\leq \gamma^{1-q}$, since $ \gamma^{j-1},(1- \epsilon  /(1-\gamma))\leq 1$. This contract also satisfies individual rationality since for $a_j=a^*(v)$, it must hold that $\mathbb{E}_{o\sim F_{\mid a_j}}[v^1(o)]\geq q+\psi(a_j)=\mathbb{E}_{o\sim F_{\mid a_j}}[t^1(v,o)]$. The inequality is immediate for $j=1$, since $\psi(a_j)=0$ and $v\in \mathcal{V}=\mathbb{R}^2_{\geq q}$. Suppose towards a contradiction that it does not hold for $j>1$, then $\sw^{a_1}(v)\geq q \geq  \mathbb{E}_{o\sim F_{\mid a_j}}[v^1(o)]-\psi(a_j)=\sw^{a_j}(v)$, in contradiction to the fact that $a^*(v)=a_j$.

\section{IIVCG Contracts: Missing Proofs from Section~\ref{sec:IIVCGclass}}

\subsection{Uniqueness of IIVCG Contracts: Proof of Theorem~\ref{thm:uniqness}}
\label{appx:holms}

To prove Theorem~\ref{thm:uniqness} we rely on a result of Holmstrom \cite{Holmstrom79}:

\begin{lemma}[Lemma $1$ in \cite{Holmstrom79}]\label{lemma:holmstrom-1}
  Let $f:[0,1]^2\to \mathbb{R}$ and $g:[0,1]\to \mathbb{R}$ satisfy:
  \begin{enumerate}
      \item $y\in \arg\max_{m\in [0,1]}f(m,y),$\label{item:holmstron-f}
      \item $y\in \arg\max_{m\in [0,1]}f(m,y)+g(m),$\label{item:holmstron-f-g}
      \item $\bigabs{\frac{\partial f(m,y)}{\partial y}}\leq K < \infty$ for all $m,y \in [0,1].$\label{item:holmstron-partial}
  \end{enumerate}
  Then $g(m)=$ constant on $[0,1].$
\end{lemma}

\begin{proof}[Proof of Theorem \ref{thm:uniqness}] 

\noindent
Define $h^{\ell}:\mathcal{V}\to\mathbb{R}$ as follows: $h^{\ell}(b)=\mathbb{E}_{o \sim F_{\mid a^*(b)}}[t^{\ell}(b,o)]+\sw^{a^*(b)}(b^{-\ell},\textbf{0})$ $\forall b\in \mathcal{V}$ and fix an arbitrary $b^{-\ell}\in \mathcal{V}^{-\ell}$. To show that~$t$ is an IIVCG contract, it suffices to show that $h^{\ell}(b^{-\ell},\cdot)$ is constant on~$b^{\ell}$. We make the following observations:
First, bidding truthfully is socially efficient, i.e.,
\begin{equation}\label{eq:uniqe-maxwel}
     v^{\ell} \in \arg\max_{b^{\ell}\in \mathcal{V}^{\ell}}\sw^{a^*(b)}(b^{-\ell},v^{\ell})\bigequationbreak {\forall v^{\ell}\in \mathcal{V}^{\ell}}.
\end{equation}
Second, since principal~$\ell$ is truthful, $v^{\ell} \in \arg\max_{b^{\ell}\in \mathcal{V}^{\ell}}\mathbb{E}_{o \sim F_{\mid a^*(b)}}[v^{\ell}(o)-t^{\ell}(b,o)] $ ${\forall v^{\ell}\in \mathcal{V}^{\ell}}.$ Replacing $\mathbb{E}_{o \sim F_{\mid a^*(b)}}[t^{\ell}(b,o)]$ with $h^{\ell}(b)-\sw^{a^*(b)}(b^{-\ell},\textbf{0})$, it holds that
\begin{equation}\label{eq:uniqe-ds}
    v^{\ell} \in \arg\max_{b^{\ell}\in \mathcal{V}^{\ell}}\sw^{a^*(b)}(b^{-\ell},v^{\ell})-h^{\ell}(b) \equationbreak {\forall v^{\ell}\in \mathcal{V}^{\ell}}.
\end{equation}

After establishing (\ref{eq:uniqe-maxwel}) and (\ref{eq:uniqe-ds}), we show that $h^{\ell}(b^{-\ell},\cdot)$ is constant on~$b^{\ell}$. Equivalently, we show that for every pair of bid profiles~$b^{\ell}, \tilde{b}^{\ell} \in \mathcal{V}^{\ell}$ $h^{\ell}(b^{-\ell},b^{\ell})=h^{\ell}(b^{-\ell},\tilde{b}^{\ell})$. To do so, define $\mathcal{V}^{\ell}(b^{\ell}, \tilde{b}^{\ell})=\{v^{\ell}(y)\mid y\in [0,1]\}$, where $v^{\ell}(y):\mathcal{O}\to \mathbb{R}$ as follows
$
 {v}^{\ell}(y;o)=y b^{\ell}(o)+(1-y)\tilde{b}^{\ell}(o) $ $\forall o \in \mathcal{O}, y\in [0,1].
$
Note that $v^{\ell}(1)=b^{\ell}$ and that $v^{\ell}(0)=\tilde{b}^{\ell}$. Thus, it suffices to show that $h^{\ell}(b^{-\ell},v^{\ell}(1))=h^{\ell}(b^{-\ell},v^{\ell}(0))$. By the assumption that $\mathcal{V}$ is convex, $\mathcal{V}^{\ell}(b^{\ell},\tilde{b}^{\ell})$ is a sub domain of $\mathcal{V}^{\ell}$. Hence, it follows from~{(\ref{eq:uniqe-maxwel})} and (\ref{eq:uniqe-ds}) that, using the notation of $a(m)=a^*(b^{-\ell},v^{\ell}(m))$,
\begin{equation}\label{eq:uniqe-maxwel-y}
     y \in \underset{m\in [0,1]}{\arg\max}\sw^{a(m)}(b^{-\ell},v^{\ell}(y)) \bigequationbreak \forall y\in [0,1]. 
\end{equation}
\begin{equation}\label{eq:uniqe-ds-y}
    y \in \underset{m\in [0,1]}{\arg\max} \sw^{a(m)}(b^{-\ell},v^{\ell}(y))-h^{\ell}(b^{-\ell},v^{\ell}(m)) \bigequationbreak \forall y\in [0,1].
\end{equation}
Define $f(m,y)=\sw^{a(m)}(b^{-\ell},v^{\ell}(y))$ and $g(m)=-h^{\ell}(b^{-\ell},v^{\ell}(m))$  $\forall m,y\in [0,1]$. With these notations, note that (\ref{eq:uniqe-maxwel-y}), (\ref{eq:uniqe-ds-y}) are exactly~\ref{item:holmstron-f},\ref{item:holmstron-f-g} in Lemma~\ref{lemma:holmstrom-1} (respectively). To apply Lemma~\ref{lemma:holmstrom-1} we show that~\ref{item:holmstron-partial} holds. Define
$
K=\arg\max_{a\in \mathcal{A}}\bigabs{{ \mathbb{E}_{o \sim F_{\mid a}}[b^{\ell}(o)-\tilde{b}^{\ell}(o)]}}.
$
Since $\mathcal{A}$ is a finite set, $K < \infty$. We show that $\bigabs{\frac{\partial f(m,y)}{\partial y}}\leq K$ for all $m,y\in [0,1]$ as follows. By definition of $f$ and of the social welfare~$\sw(\cdot)$,
\begin{equation}\label{eq:holms-partial}
\biggabs{\frac{\partial f(m,y)}{\partial y}}{}=\biggabs{\frac{\partial \sw^{a(m)}(b^{-\ell},v^{\ell}(y))}{\partial y}}=\biggabs{\frac{\partial \mathbb{E}_{o \sim F_{\mid a(m)}}[v^{\ell}(y,o)]}{\partial y}}
\end{equation}
By definition of $v^{\ell}(y,o)$,  
\begin{equation}\label{eq:holms-vsy}
    \mathbb{E}_{o \sim F_{\mid a(m)}}[v^{\ell}(y,o)]=y\cdot \mathbb{E}_{o \sim F_{\mid a(m)}}[b^{\ell}(o)]+(1-y)\cdot \mathbb{E}_{o \sim F_{\mid a(m)}}[\tilde{b}^{\ell}(o)].
\end{equation}
According to (\ref{eq:holms-partial}) and (\ref{eq:holms-vsy}) and definition of~$K$,
\begin{equation*}
\biggabs{\frac{\partial \mathbb{E}_{o \sim F_{\mid a(m)}}[v^{\ell}(y,o)]}{\partial y}}= \bigabs{ { \mathbb{E}_{o \sim F_{\mid a^*(m)}}[b^{\ell}(o)-\tilde{b}^{\ell}(o)]}}\leq K.
\end{equation*}
By Lemma~\ref{lemma:holmstrom-1}, $g(m)$ is constant on $m$. That is,  $h^{\ell}(b^{-\ell},v^{\ell}(m))$ is constant on $m$. We conclude that $h^{\ell}(b^{-\ell},v^{\ell}(1))=h^{\ell}(b^{-\ell},v^{\ell}(0))$ and the proof is complete. 
\end{proof}

\subsection{Characterization of IIVCG: Proof of Lemma~\ref{lemma:structure}}
\label{appx:structure}

\begin{proof}[Proof of Lemma \ref{lemma:structure}]
We start with the forward direction. Let~$t$ be an IIVCG contract. 
Note that ${{t^{\ell}}:  \mathcal{V}\times \mathcal{O}\rightarrow \mathbb{R}}$ can be re-composed as the sum of two functions; ${c^{\ell}: \mathcal{V} \rightarrow \mathbb{R},}$ that depends only on~$b$, and ${g^{\ell}: \mathcal{V}\times \mathcal{O} \rightarrow \mathbb{R}}$ that depends on $b$ as well as on the realised outcome, i.e., $t^{\ell}(b,o)=c^{\ell}(b)+g^{\ell}(b,o)$. With these notations, for every principal~$\ell$ and every bid profile~$b$,
\begin{equation}\label{eq:exp-c}
\mathbb{E}_{o \sim F_{\mid a^*(b)}}[t^{\ell}(b,o)]=c^{\ell}(b)+ \mathbb{E}_{o \sim F_{\mid a^*(b)}}[g^{\ell}(b,o)].
\end{equation}
By IIVCG Definition, there exist $\{h^{\ell}\}$ such that
\begin{equation}\label{eq:exp-h}
        \mathbb{E}_{o \sim F_{\mid a^*(b)}}[ t^{\ell}(b,o)]=h^{\ell}(b^{-\ell})-\sw^{a^*(b)}(b^{-\ell},\textbf{0}), 
\end{equation}
where $h^{\ell}$ does not depend on $b^{\ell}$. 
Equalizing (\ref{eq:exp-c}), (\ref{eq:exp-h}) we get
$c^{\ell}(b)=h^{\ell}(b^{-\ell})-\sw^{a^*(b)}(b^{-\ell},g^{\ell}(b,o)).
$
Since $t^{\ell}(b,o)=c^{\ell}(b)+g^{\ell}(b,o)$, $t^{\ell}(b,o)=h^{\ell}(b^{-\ell})-\sw^{a^*(b)}(b^{-\ell},g^{\ell}(b,o))+g^{\ell}(b,o).$
By IIVCG contract definition, the agent maximizes the (declared) social welfare, i.e., $x^*(b)=a^*(b)$.
Thus, 
$
a^*(b)\in \arg\max_{a\in \mathcal{A}}\mathbb{E}_{o \sim F_{\mid a}}[\sum_{{\ell \in [n]}} t^{\ell}(b,o)]-\psi(a).
$
The only outcome-depended term in $t^{\ell}(b,o)$ is $g^{\ell}(b,o)$.\footnote{If $h^{\ell}(b^{-\ell})$ depends on the realised outcome, then it would also depend on principal $\ell$'s bid (since $\ell$'s bid influences the realised outcome indirectly by influencing the chosen action).} Hence,
\begin{equation}\label{eq:max-g}
a^*(b)\in \arg\max_{a\in \mathcal{A}}\mathbb{E}_{o \sim F_{\mid a}}[\sum_{{\ell \in [n]}}g^{\ell}(b,o)]-\psi(a).
\end{equation} Define $w^{\ell} : \mathcal{O}\to \mathbb{R}^m_{\geq 0}$ as follows:
$w^{\ell}(o) = g^{\ell}(b,o)-\min_{o\in \mathcal{O}}g^{\ell}(b,o)$ $\forall o\in \mathcal{O}.
$
Hence, $
        t^{\ell}(b,o)=h^{\ell}(b^{-\ell})-\sw^{a^*(b)}(b^{\ell},w^{\ell})+w^{\ell}(o),$
where $\sum_{{\ell \in [n]}}w^{\ell} \in \mathcal{L}_{a^*{(b)}}$ by~\eqref{eq:max-g} and the definition of $\mathcal{L}_a$.

For the backward direction, let~$t$ be a contract as specified in Lemma~\ref{lemma:structure}.
Note that the only outcome dependent term in $t^{\ell}(b,o)$ is $w^{\ell}(o)$. Thus,
\begin{equation}\label{eq:t-and-w-proof:structure}
\arg\max_{a\in \mathcal{A}}\mathbb{E}_{o \sim F_{\mid a}}[\sum_{{\ell \in [n]}} t^{\ell}(b,o)]-\psi(a)=\arg\max_{a\in \mathcal{A}}\mathbb{E}_{o \sim F_{\mid a}}[\sum_{{\ell \in [n]}} w^{\ell}(o)]-\psi(a).
\end{equation} Since, $\sum_{{\ell \in [n]}} w^{\ell}\in \mathcal{L}_{a^*{(b)}}$, $a^*(b)\in \arg\max_{a\in \mathcal{A}}\mathbb{E}_{o \sim F_{\mid }}[\sum_{{\ell \in [n]}} w^{\ell}(o)]-\psi(a).$
Thus, and according to~(\ref{eq:t-and-w-proof:structure}) we conclude that $a^*(b)$ maximizes the agent's utility. Furthermore, Property~\ref{def:IIVCG-item:hs} in the definition of IIVCG (Definition~\ref{def:IIVCG}) is also met. since principal $\ell$'s expected payment is $
\expbest{t^{\ell}(b,o)}=h^{\ell}(b^{-\ell})-\sw^{a^*(b)}(b^{\ell},w^{\ell}) + \expbest{w^{\ell}(o)}.
$
After rearranging the above $\expbest{t^{\ell}(b,o)}=h^{\ell}(b^{-\ell})-\sw^{a^*(b)}(b^{\ell},\textbf{0})$ as desired.
\end{proof}

\subsection{A Trade-off between LL and IR: Proof of Theorem~\ref{thm:notAlwaysVCG}}
\label{appx:trade}

Here, we formally prove Theorem~\ref{thm:notAlwaysVCG}. 
\begin{example}\label{ex:trade-off}
There are two actions~$a_1,a_2$ and two possible outcomes~$o_1,o_2$. The distributions imposed by actions are:~$p_{o\sim F_{\mid a_1}}(o_1)=\frac{1}{2},$ $p_{o\sim F_{\mid a_2}}(o_1)=1.$ There is one principal with valuation domain~$\mathcal{V}^{1}=\mathbb{R}^2_{\geq 0}$.
The cost for each action is: $\psi(a_1)=0,$ $\psi(a_2)=\epsilon.$
\end{example}

\begin{proof}[Proof of Theorem~\ref{thm:notAlwaysVCG}]
Consider the common agency setting in Example~\ref{ex:trade-off} and the valuation profile~$v^{1}=(3\epsilon,0).$ Note that $\sw^{a_1}(v)=1.5\epsilon$, and $\sw^{a_2}(v)=2\epsilon$. Thus, the socially efficient action is~$a_2$. Suppose toward a contradiction that there exists an IIVCG contract~$t$ that satisfies LL and IR. Since~$t$ is an IIVCG contract, in a dominant strategy equilibrium the principal is truthful, and the agent takes the socially efficient action, i.e.,~$a_2$.
Since the agent maximizes the sum of payments minus cost,
it must hold that 
$
\sum_{\ell \in [n]} t^{\ell}(b,o_1)-\epsilon\geq \sum_{o\in\mathcal{O}}\frac{1}{2}\cdot\sum_{\ell \in [n]} t^{\ell}(b,o).
$
Thus, 
$
t^{1}(b,o_1)- t^{1}(b,o_2) \geq 2\epsilon.
$
By the assumption that $t$ satisfies LL, $t^{1}(b,o_2)\geq 0$, so, $t^{{1}}(b,o_1) \geq 2\epsilon$. That is, $t^{1}(b,o_1)=\mathbb{E}_{o \sim F_{\mid a_2}}[t^{1}(b,o)]=h^{1}(b^{-1})-\sw^{a_2}(b^{-1},\textbf{0})\geq 2\epsilon$. Since $\sw^{a_2}(b^{-1},\textbf{0})=-\epsilon$, we conclude that~$h^{1}(b^{-1})\geq \epsilon$.
Now, consider the following valuation profile: $\tilde{v}^{1}=(0,0)$. Again, in a dominant strategy equilibrium the principal is truthful $\tilde{b}=\tilde{v}$ and the agent takes the socially efficient action, which is now $a_1$.
By definition of IIVCG, $h^{1}$ is independent of $b^1$. Thus, $h^{{1}}(b^{-{1}})=h^{{1}}(\tilde{b}^{-{1}})\geq \epsilon$.
Recall that $\mathbb{E}_{o \sim F_{\mid a_1}}[t^{1}(\tilde{b},o)]=h^{1}(\tilde{b}^{-1})-\sw^{a_1}(\tilde{b}^{-1},\textbf{0})$, where $\sw^{a_1}(\tilde{b}^{-1},\textbf{0})=0.$ Thus,
$\mathbb{E}_{o \sim F_{\mid a_1}}[t^{1}(\tilde{b},o)]=h^{1}(\tilde{b}^{-1})\geq \epsilon$, while $\mathbb{E}_{o \sim F_{\mid a_2}}[\tilde{v}^{1}(o)]=0$. This is a contradiction to the fact that $t$ satisfies IR.
\end{proof}



\end{document}